\documentclass[journal]{IEEEtran}

\usepackage{amssymb}
\usepackage{amsmath}

\usepackage{amsfonts}
\usepackage{gensymb}
\usepackage{algorithmic}
\usepackage{algorithm}
\usepackage{array}
\usepackage[caption=false,font=scriptsize,labelfont=sf,textfont=sf]{subfig}
\usepackage{textcomp}
\usepackage{stfloats}
\usepackage{url}
\usepackage[hidelinks]{hyperref}
\usepackage{orcidlink}
\usepackage{verbatim}
\usepackage{graphicx, color, xcolor}
\usepackage{tabularx}
\usepackage{multirow}
\usepackage{booktabs} 
\usepackage{bm} 
\usepackage{wrapfig}
\DeclareMathOperator*{\argmax}{\textit{arg max}}
\DeclareMathOperator*{\argmin}{\textit{arg min}}
\usepackage{makecell}

\begin{document}
\title{Exploiting Age of Information in Network Digital Twins for AI-driven Real-Time Link Blockage Detection}
\author{
    Michele Zhu\orcidlink{0009-0003-7448-559X},
    Francesco Lisalata\orcidlink{0000-0002-6725-3606},
    Silvia Mura\orcidlink{0000-0002-0207-5730},\\
    Lorenzo Cazzella\orcidlink{0000-0002-8484-3536},
    Damiano Badini,
    Umberto Spagnolini\orcidlink{0000-0002-7047-2455}\\
    \thanks{M. Zhu, S. Mura, F. Lisalata, L. Cazzella, and U. Spagnolini are with Politecnico di Milano. D. Badini is with Huawei Technologies Italia.}
}

\maketitle
\begin{abstract}
The Line-of-Sight (LoS) identification is crucial to ensure reliable high-frequency communication links, especially those vulnerable to blockages. Network Digital Twins and Artificial Intelligence are key technologies enabling blockage detection (LoS identification) for high-frequency wireless systems, e.g., 6$>$GHz. In this work, we enhance Network Digital Twins by incorporating Age of Information (AoI) metrics, a quantification of status update freshness, enabling reliable real-time blockage detection (LoS identification) in dynamic wireless environments. By integrating raytracing techniques, we automate large-scale collection and labeling of channel data, specifically tailored to the evolving conditions of the environment. The introduced AoI is integrated with the loss function to prioritize more recent information to fine-tune deep learning models in case of performance degradation (model drift).
The effectiveness of the proposed solution is demonstrated in realistic urban simulations, highlighting the trade-off between input resolution, computational cost, and model performance. A resolution reduction of $4\times8$ from an original channel sample size of $(32, 1024)$ along the angle and subcarrier dimension results in a computational speedup of $32$ times. 
The proposed fine-tuning successfully mitigates performance degradation while requiring only $1\%$ of the available data samples, enabling automated and fast mitigation of model drifts. 
\end{abstract}




\section{Introduction}
\label{sect:introduction}

The integration between Network Digital Twins (NDTs) and Artificial Intelligence (AI) will play a pivotal role in sixth-generation (6G) wireless systems. A key requirement in data-driven AI, i.e., Machine Learning (ML), is the availability of high-quality, large-scale training datasets.
Yet, the compilation and annotation of these extensive datasets typically require a human-in-the-loop methodology, which is prone to errors and incurs substantial costs. Thus, it is inadequate to meet the demands of 6G networks. In contrast, NDTs can collect high-quality data and automate the labeling process, accelerating the ML training workflow~\cite{Lv2022}. However, the dynamic nature of wireless environments can quickly deteriorate the relevance of the collected data. 

The Age of Information (AoI) metric describes the timeliness of a monitor's knowledge of a process. Kaul et al.~\cite{5984917} introduced the metric for Ultra Reliable Low Latency (URLL) communication in safety messaging. By integrating AoI mechanisms, less recent data can be automatically excluded from the \textit{machine learning} model training process. Allowing the model to be trained on the current context and relevant information, ensuring their performance and reliability in ever-evolving network conditions.

As networks become increasingly complex, AI algorithms have demonstrated excellent performance in solving challenging tasks. Some examples, though not exhaustive, include network management~\cite{5415567}, spectrum optimization~\cite{7470268}, spectrum sharing~\cite{10643670}, and sensing~\cite{10.1007/s11277-023-10630-x}.
Moreover, ML models can quickly and accurately determine the blockage condition between the transmitter and receiver, but researchers have not yet explored strategies to guarantee their performance over time. In high-frequency bands, i.e., Frequency Range 2 (FR2) and Frequency Range 3 (FR3), the Non-Line-of-sight (NLoS) condition can severely affect network coverage and reliability~\cite{10459211, 6515173}. Blockage detection enhances communication performance by improving network reliability and facilitates advanced sensing capabilities in fifth-generation (5G) and emerging 6G networks~\cite{linsalata2022map, 9364875}.

This manuscript will focus on improving the data acquisition process by integrating NDTs for link blockage identification. In addition, will introduce an AoI-aware loss function, adapting the model training process to prioritize recent data, ensuring robust ML models in non-stationary environments such as wireless communication.

\subsection{Related Works}
This subsection provides an overview of the relevant literature in the domain, highlighting key research contributions, existing methodologies, and state-of-the-art approaches. We aim to establish a clear understanding of the previous progress and lay the foundation for the development of our solution.

FR2 and Massive Multiple Input Output - Orthogonal Frequency Division Multiplexing (MIMO-OFDM) form the foundation of 5G New Radio (NR) standards~\cite{3gpp_38855}. In MIMO systems, each antenna is digitally oriented to a specific mobile device, greatly increasing signal quality and enabling higher data rates~\cite{7064850}. OFDM is a widespread digital modulation technique in wireless communications. The signal is split into multiple narrow-band subcarriers, and symbol interference is mitigated using the cyclic prefix~\cite{9559117}. Measurement campaigns have demonstrated the difference between Line-of-Sight (LoS) and NLoS path loss at 28 and 38 GHz~\cite{6515173}, the absence of the LoS path significantly affects signal quality.

Existing \textit{blockage detection} methodologies can be categorized into model-based and Machine Learning (ML) approaches:
\textit{i)} model-based approach rely on well-defined environment models; examples include decision-theoretic frameworks based on range measurements~\cite{686556}, correlation metrics across time, space, and frequency~\cite{8886022}, Rician K-factor~\cite{9552572}.
On the other hand, \textit{ii)} data-driven (ML) approaches leverage large data collection of Channel State Information (CSI) measurements to learn the relationship between blockage condition and the measurements. They can be further divided into: \textit{iia)} Statistical Machine Learning (SML), i.e., methods such as Support Vector Machines (SVMs)~\cite{8968748}, Random Forests (RFs)~\cite{10359439}, and Gaussian Processes (GPs)~\cite{8322223} that summarize the measurements into their descriptive statistics/patterns before learning the relationship between prediction labels and data; \textit{iib)} Deep Learning (DL)~\cite{9264213}, i.e., methods that process the raw CSI to directly learn the patterns that distinguish the Line-of-Sight (LoS) condition from NLoS condition. Data-driven approaches outperform and offer greater flexibility than model-based approaches. Nevertheless, they rely on the availability of high-quality, large-quantity data samples.

\textit{Radio Frequency Fingerprinting} (RFF) has gained attention as a method for collecting device samples. RFF relies on the assumption of a static communication channel environment. The data collection process for RFF involves generating brief intermittent snapshots of the channel impulse response (CIR) of the wireless device over a long time interval, such as several days~\cite{9682024, 8968748}. RFF has been applied in device identification~\cite{9685067}, intrusion detection~\cite{8795319}, modulation classification~\cite{10.1007/978-3-319-73564-1_37}, and localization~\cite{9364875}. However, RFF overlooks the non-stationarity of wireless environments and scalability issues in large-scale wireless network deployments, making it unsuitable in dynamic environments.

\textit{Network Digital Twins} are envisioned to revolutionize 6G wireless systems by integrating simulations and models with real-world sensory capabilities, allowing them to capture the entire evolution of the system.
Today we have precise 3D modeling capabilities with different types of sensor: camera~\cite{rs5126880}, LIDAR~\cite{ZHAO2023101974}, and radar~\cite{9973561}. Future connected autonomous vehicles will mount the aforementioned sensors, providing sensing capabilities of dynamic environments~\cite{IGNATIOUS2022736}.
Alkhateeb et al.~\cite{10198573} propose a physical layer NDT, enabled by high-accuracy 3D digital maps and multimodal sensing. The NDT-enabled framework in~\cite{10757947} demonstrates improved performance in high-frequency V2X handover tasks. The estimation of the wireless channel with NDT is shown in~\cite{delmoro2025bayesianemdigitaltwins}.
Those NDT models are characterized by their deterministic channel modeling: \textit{raytracing}.
The raytracing method is based on the geometric computation of propagation path given a 3D model of the environment, and the propagation effects are obtained by numerically solving the Maxwell equations given the electromagnetic (EM) properties of the environment and transmitted signal.
Recent advances in raytracing modeling, propagation path computation, and hardware acceleration have shown promising results in obtaining near-real-time channel modeling~\cite{zhu2024realtime, pegurri2025digitalnetworktwinsintegrating}. 

\textit{Age of Information} is a metric that quantifies the freshness of our knowledge concerning the status of a remote system~\cite{8187436}. It represents the duration between the current timestamp at the receiver end and the timestamp of the status update when it was generated at the transmitter. Optimizing communication systems based on the AoI metric is essential for URLL systems, such as autonomous driving and smart factories, where stringent requirements on information freshness are imperative to ensure the required functionality. As demonstrated by the authors in~\cite{6195689}, minimizing AoI requires a balance between the update rate and the congestion of the network. Furthermore, the implications of channel utilization on AoI optimization are explored in~\cite{8445979, 9377549}. The importance of information freshness has a wide range of applications, such as data warehouses, communication networks, and control systems. However, its application to NDT-empowered 6G networks has never been approached so far, and it is the topic addresses in this paper.

\subsection{Contributions}
This paper is contextualized in the future NDT-enabled wireless system and its integration with ML models. The main contributions of the paper are summarized as follows.
\begin{itemize}
    \item Proposes the NDT data collection process to automate site-specific \textit{data-driven} model deployment, evaluating SVM, RF, and DL implementations. 
    \item For DL models, the input resolution reduction and data augmentation have been analyzed, showing a computational speedup of 32 times and improvement of approximately $5\%$ for SNR $<$ 10 dB, compared to literature.
    \item Introduces the AoI metric into the collected data samples and the AoI-aware loss function to fine-tune DL models. The AoI-ware loss formulation weights model optimization toward more recent data samples, decreasing the number of samples needed to fine-tune the model up to $1\%$ of the available data and achieving accuracy over $98\%$.
    \item Simulations are performed in a static wireless environment, where we compare with the latest method~\cite{8968748, 10359439}, and in a dynamic wireless environment, i.e., moving scattering objects.
\end{itemize}

\subsection{Organization and Notation}
\textbf{Organization.} The remainder of this paper is organized as follows. Sec.~\ref{sec: MIMO-OFDM and ML inputs} presents the components of an NDT for the physical layer, the considered MIMO-OFDM system, and the ML inputs. The following Sec.~\ref{sec: blockage det} details the proposed solution, in particular, the role of NDT in an automated data collection is discussed, the AoI metric and AoI-aware loss formulation are introduced to fine-tune DL models.
Sec.~\ref{sec: experimental setup} describes the scenario considered, the collected data, and the training and evaluation of the ML model. Sec.~\ref{sec: num analysis} analyzes the experimental results, while Sec.~\ref{sec: discussion} contains a critical discussion of the proposed solution. The conclusions are drawn in Sec.~\ref{sec: conclusion}.

\textbf{Notation.} $(\cdot)^*$, $(\cdot)^T$ and $(\cdot)^H$ denote complex conjugate, transpose, and Hermitian operations. The bold lowercase and uppercase letters represent vectors and matrices, respectively. The Kronecker product is $\otimes$ and the Hadamard product is $\odot$. 
The subscript $i$ will be used to represent an element in dataset $\mathcal{N}$, where $\mathcal{N}$ is used to denote an unspecified but bounded dataset.

\section{System Model} \label{sec: MIMO-OFDM and ML inputs}
In this section, we describe the wireless channel model under consideration and the transformation applied to channel measurements.

\subsection{Network Digital Twin for the Physical Layer}
A Network Digital Twin for the physical layer can be conceptualized as two distinct components:
\begin{itemize}
    \item[\textit{i)}] the \textit{communication environment};
    \item[\textit{ii)}] the \textit{raytracer}.
\end{itemize}
The environment \textit{(i)} is represented by a 3D model that encapsulates the position, orientation, velocity, and shapes of the object within the environment. These objects include Base Stations (BSs), User Equipments (UEs), and environmental scatterers. The electromagnetic properties of the objects are incorporated into the material specifications of the 3D models. Raytracing \textit{(ii)} simulates the multipath propagation between each transmitter pair that accounts for reflections, diffractions, and scattering. The propagation paths can be used to synthesize the communication channel.

\subsection{Wireless Channel  Model}

The cellular wireless system under consideration comprises a BS mounted with a Uniform Planar Array (UPA) consisting of $N_h \times N_v$ antennas, operating at a carrier frequency of $f_c$ and having a bandwidth of $B$. In this configuration, UE establishes an uplink communication with the BS, using an omnidirectional antenna. The transmission of signals is performed through an OFDM waveform characterized by a sampling interval of $T_s$ and $N_c$ subcarriers.

The Channel Frequency Response (CFR) at the $l$th OFDM subcarrier, the link between a UE and the BS, is modeled as follows:
\begin{equation}
    \mathbf{h} [l] = \sum^{N_p}_{p=1} \alpha_{p} \mathbf{e} (\vartheta_{p}, \varphi_{p}) \in \mathbb{C}^{N_v N_h \times 1},
\end{equation}
where $\alpha_{p} = a_{p} e^{-j2\pi \tau_{p} f_l}$ is the complex path gain, with $a_{p}$ denoting the path amplitude, $\tau_p$ the path delay, the vector $\mathbf{e}(\vartheta_{p}, \varphi_{p})$ represents the array response \cite{9364875}, and $N_p$ denotes the number of multipath components encountered in the communication.

The Space-Frequency domain Channel Response Matrix (SFCRM) is the collection of CFRs over the $N_c$ subcarriers:
\begin{equation} \label{eq: channel frequency response}
    \mathbf{H} \triangleq \left[ \mathbf{h}\left[0\right] , \mathbf{h}\left[1\right], ..., \mathbf{h}\left[N_c -1\right]\right]\in \mathbb{C}^{N_v N_h \times N_c},
\end{equation}
The received signal $\mathbf{r}[l]  \in \mathbb{C}^{N_v N_h \times 1}$ is
\begin{equation}\label{eq: AWGN}
    \mathbf{r}[l] = \mathbf{h}[l] {s}[l] + \mathbf{z}[l],
\end{equation}
where ${s}[l]$ represents the transmitted symbol data such that $\mathbb{E}[{s}[l] {s}[l']^*] = \sigma_s^2 \delta(l-l^\prime)$ and $\mathbf{z}[l] \sim \mathcal{CN}(0, \sigma_{z}^2\, \mathbf{I}) $ is the AWGN. 

\subsection{Feature engineering for Statistical Machine Learning} \label{subsec: sml-features}
Feature engineering is the process of transforming raw data (channel measurements) into measurable properties (features). The process encompasses the identification, selection, and creation of significant features that enable the model to discern relevant patterns in the data. The choice of informative and discriminating features is crucial for SML algorithms. The wireless channel between the BS and the UE is completely characterized by its multipath propagation components. Each propagation path is described by its power, delay, azimuth, and elevation angle. Real-world multipaths are derived with high-resolution parameter estimations~\cite{doi:https://doi.org/10.1002/9781119294016.ch7} from channel measurements. Raytracing directly provides multipath estimates based on the 3D model of the environment.
From real-world data or NDT, we define the following features:
\begin{itemize}
    \item \textit{Received Signal Strength} ($P_{RSS}$), quantifies the total received signal power, assuming a unitary transmitted power, and is defined as follows:
    \begin{equation}
        P_{RSS} = \sum_{p=1}^{N_p} |\alpha_{p}|^2,
    \end{equation}
    where $N_p$ represents the number of paths identified by the NDT, or measured from the channel estimates. The LoS condition demonstrates greater total power than the NLoS condition, which involves a longer travel distance.
    \item \textit{Maximum Received Power} ($P_{max}$), denotes the maximum power over all paths, defined as:
    \begin{equation}
        P_{max} = \max |\alpha_{p}|^2,
    \end{equation}
    The LoS path typically exhibits greater power than the NLoS paths.

    \item \textit{Root Mean Squared Delay Spread} ($\tau_{rms}$), constitutes a fundamental metric in wireless communication, quantifying the temporal dispersion of a transmitted signal as a consequence of multipath propagation. It can be expressed as:
    \begin{equation}
        \tau_{rms} = \sqrt{\sum^{N_p}_{p=1} \eta_p (\tau_p - \overline{\tau}_p)},
    \end{equation}
    where $\overline{\tau} = \sum^{N_p}_p \eta_p \tau_p$ is the mean delay and
    \begin{equation}
        \eta_p = \frac{|\alpha_p|^2}{\sum^{N_p}_{p=1} |\alpha_p|^2} = \frac{|\alpha_p|^2}{P_{RSS}} \nonumber
    \end{equation}

    is the normalized power of the p$th$ path. In the NLoS channel, $\tau_{rms}$ generally exhibits a greater delay spread.

    \item \textit{Rise-Time} ($\Delta\tau$), represents the time interval between the strongest path ($\tau_{max}$) and the weakest path ($\tau_{min}$), it is written as:
    \begin{equation}
        \Delta\tau = \tau_{max} - \tau_{min}.
    \end{equation}
    The rise time $\Delta\tau$ is typically more pronounced under NLoS conditions due to the increased delays.
    \item \textit{Angle of Arrival Spread of azimuth and elevation} ($\vartheta_{rms}$,$\varphi_{rms}$), measures the dispersion of angles in the respective directions. The angular spread is obtained as:
    \begin{equation}
         \begin{cases}
             \vartheta_{rms} = \sum^{N_p}_{p=1} \eta_p \vartheta_p \\
             \varphi_{rms} = \sum^{N_p}_{p=1} \eta_p \varphi_p
         \end{cases}
    \end{equation}
    with $\vartheta_{p}$ referring to the azimuth angle and $\varphi_{p}$ elevation angle of the p$th$ path. The angle spread of the LoS condition is generally smaller than that of the NLoS condition.
\end{itemize}
\bigskip
The input feature vector for SML algorithms is written as:
\begin{equation}
    \bm{\psi} = [P_{RSS}, P_{max}, \tau_{rms}, \Delta_{\tau}, \vartheta_{rms}, \varphi_{rsm}]^T.
\end{equation}
It is important to remark that in the context of \textit{digital twin}, inevitable discrepancies may arise between synthetic and real-world multipath propagation. In pratical scenarios, antennas are typically unable to resolve individual paths that arrive as a cluster of closely spaced rays. Particularly at high frequencies, the number of distinguishable multipath components is generally limited to $N_p \leq 6$~\cite{6834753}.

\subsection{Angle delay power profile for Deep Learning} \label{subsec: adcpm-def}
Unlike \textit{Statistical Machine Learning} (SML) methodologies, DL architectures are capable of processing raw input data to perform feature extraction. Although the complex-valued SFCRM representation is compatible with specialized DL architectures, its integration presents considerable difficulties in terms of architectural design and parameter optimization. It is essential to convert it into an appropriate format, the angle delay power profile, known as the Angle Delay Channel Power Matrix (ADCPM), defined as:
\begin{equation}\label{eq:adcpm}
    \mathbf{P} \triangleq \mathbb{E}\left[\mathbf{G} \odot \mathbf{G}^*\right] \in \mathbb{R}^{N_v N_h\times N_c}.
\end{equation}
Where $\mathbf{G} \in \mathbb{C}^{N_v N_h \times N_c}$ is the angle delay channel response matrix obtained over array ($\mathbf{V}_v$, $\mathbf{V}_h$) and time ($\mathbf{F}$)~\cite{9364875}:
\begin{equation} 
    \mathbf{G} \triangleq \frac{1}{\sqrt{N_v N_h N_c}} \left( \mathbf{V}_v \otimes \mathbf{V}_h \right)^H \mathbf{H} \mathbf{F}^*,
\end{equation}
with $\mathbf{V}_v \in \mathbb{C}^{N_v \times N_v}$ , $\mathbf{V}_h \in \mathbb{C}^{N_h\times N_h}$ and $\mathbf{F} \in \mathbb{C}^{N_c \times N_h N_v}$ DFT matrices with elements defined as:
\begin{equation}
    \begin{cases}
        [\mathbf{V}_v]_{a,b} = \frac{1}{\sqrt{N_v}} e^{-j2\pi \frac{a(b-N_v/2)}{N_v}} \\
        [\mathbf{V}_h]_{c,d} = \frac{1}{\sqrt{N_h}} e^{-j2\pi \frac{c(d-N_h/2)}{N_h}} \\
        [\mathbf{F}]_{f,g} = \frac{1}{\sqrt{N_c}}e^{-j2\pi \frac{fg}{N_c}}
    \end{cases}
\end{equation}

\section{Proposed Blockage Detection Method} \label{sec: blockage det}

This section presents the problem formulation for blockage detection within a supervised learning framework. It describes the ML models used in the study, and the NDT data collection process. Lastly, the AoI metric in the NDT data and DL fine-tuning is introduced.

\subsection{Problem formulation}
A supervised machine learning framework is considered for the task of blockage detection. 
The goal is to determine the associated blockage condition $y\in \{0,1\}$, where $y=1$ indicates the channel in the NLoS condition and $y=0$ indicates the channel in the LoS condition, following the input-output relationship
\begin{equation}
    \hat{y} = f(\mathbf{x}; \bm{\zeta}(\mathcal{N})).
\end{equation}
Here $f(\cdot)$ is a non-linear function, $\bm{\zeta}(\mathcal{N})$ are the model parameters learned from the dataset $\mathcal{N}$, and $\hat{y}$ is the predicted blockage condition. 
In detail: 
\begin{itemize}
    \item $\mathbf{x}=\bm{\psi}$ in Sec.~\ref{subsec: sml-features} if we consider SML methods;
    \item $\mathbf{x}=\mathbf{P}$ in Sec.~\ref{subsec: adcpm-def} if we consider DL methods,
\end{itemize}
as specified below.

\subsection{Statistical Machine Learning Methods}
The following two \textit{Statistical Machine Learning} algorithms are analyzed: \textit{i)} Support Vector Machines that distinguish two classes by finding the optimal hyperplane that maximizes the margins between the closest data points of opposite classes.
Wherever the kernel function is the Radial Basis Function is referred to as SVM RBF~\cite{8968748};
\textit{ii)} Random Forests combine the outputs of multiple decision trees to reach a single decision.
In particular the Classification and Regression Trees-Random Forest (CART-RF) method~\cite{10359439} is examined. Each branching criteria of the decision tree is performed according to the Gini index~\cite{10.5555/1162264}. The decision tree output aggregation is performed according to a majority vote.

\subsection{Deep Learning Methods} \label{subsec: DL model}
Training a DL model means optimizing the model parameters $\bm{\zeta}$ for a given dataset $\mathcal{N}$, with $|\mathcal{N}| = N$. Considering a probabilistic discriminative model, the estimate by the maximum likelihood is:
\begin{equation} \label{eq: maximum-likelihood}
    \bm{\hat{\zeta}} (\mathcal{N}) = \argmax_{\bm{\zeta}} \prod^{N}_{i=1}p(y_i| f(\mathbf{P}_i; \bm{\zeta})),
\end{equation}
where $p(y_i|f(\mathbf{P}_i; \bm{\zeta}))$ represents the parametric probability, $\mathbf{P}_i$ is the ADCPM defined in Eq.~\eqref{eq:adcpm}. The index $i$ indicates the i$th$ channel measurement in the dataset $\mathcal{N}$.
The conditional probability pertaining to a binary classification task can be expressed as: $p(y_i|f(\mathbf{P}_i; \bm{\zeta})) = \hat{p}_{y_i} \delta[y_i - 1] + (1-\hat{p}_{y_i}) \delta[y_i]$. In this context, $\hat{p}_{y_i}$ denotes the prediction of the model for the positive class, contingent upon the i\textit{th} input sample.
Finally, the optimization~\eqref{eq: maximum-likelihood} becomes equivalently:
\begin{align}\label{eq: loss_bce}
    \bm{\hat{\zeta}} (\mathcal{N}) &= \argmin_{\bm{\zeta}} \sum^{N}_{i=1} -\left[y_i log (\hat{p}_{y_i}) + (1-y_i) log (1-\hat{p}_{y_i}) \right] \\
    &= \argmin_{\bm{\zeta}} \mathcal{L}(\mathcal{N}; \bm{\zeta}), \nonumber
\end{align}
where $\mathcal{L} \left( \mathcal{N}; \bm{\zeta} \right)$ denotes the loss function. In the following subsections, we will analyze a practical approach to reduce the computational cost of the DL model for blockage detection and AoI for DL.

\subsection{Network Digital Twins data collection}
Conducting measurement campaigns that capture all possible wireless configurations is impractical, and mitigating ML model degradation (model drift) requires costly and time-consuming re-measurements.
Considering a wireless system where CSI is continuously collected as: \textit{i)} crowd-sourced samples tagger with UE location, the label is provided by computing the direct path between the BS and the UE; \textit{ii)} NDT synthetic samples for underrepresented distributions. 
This integration of NDT and crowd-sourcing enables fully automated data collection, significantly reducing human intervention costs and accelerating downstream data-driven deployments. However, unbounded data collection leads to increasing dataset sizes, causing longer training times. The AoI is introduced into the NDT framework to prune the dataset and weight the optimization of model parameters toward more recent data samples.

\subsection{Age of Information for Deep Learning} \label{subsec: aoi-learning}

\begin{figure*} [!tb]
    \centering
    \subfloat
    {\includegraphics[width = 0.95\linewidth]{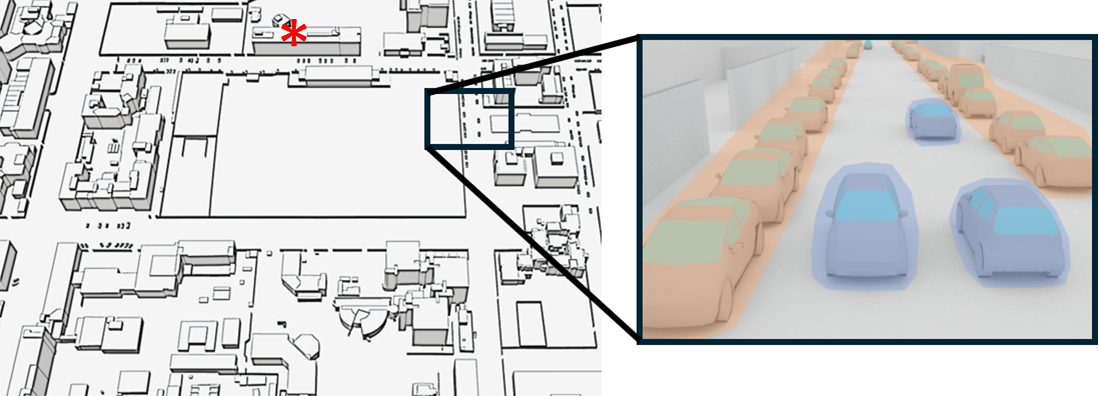}}

    \caption{Bird-eye view of the considered city section $670\text{m}\times 550\text{m}$ with parked vehicles, the red asterisk indicates the position of the Base Station. The zoom-in shows a snapshot of the simulation with dynamic wireless environment, where in orange are the parked vehicles in the scene and in blue the moving vehicles.}
    \label{fig: NDT-milan}
\end{figure*}

The inherent dynamicity of wireless environments will introduce novel patterns absent from historical data, leading to potential model drift~\cite{Vela2022}. NDTs play a pivotal role in mitigating these issues by serving as monitors to detect potential model drifts. Upon detection, corrective actions such as model fine-tuning can be applied, i.e., model parameter optimization from a previous non-random state.
We associate to a collected data samples $\mathbf{P}_i$ the concept of freshness $\Delta_i (t)$, i.e. AoI metric, as:
\begin{equation}
    \Delta_i(t) = t - u_i(t),
\end{equation}
where $t \geq u_i(t)$ is the current system timestamp and $u_i(t)$ is the timestamp when the data sample has been generated. 
The function $h(\cdot)$ excludes old data samples and is defined as follows:
\begin{equation}
    h(\mathbf{P}_i, t) = 
    \begin{cases}
        \mathbf{P}_i & \text{if } e^{-\gamma\Delta_i (t)} > \Gamma \\
        0 & \text{else.}
    \end{cases}
\end{equation}
Where $\Gamma$ is a constant denoted as the age violation threshold, the exponential decay function $e^{-\gamma\Delta_i(t)}$ is denoted as the decay factor, and $\gamma>0$ is the constant exponential decay rate. 
By reducing the number of samples, we reduce the size of the datasets, thus obtaining faster training times.

The loss function in Eq.~\eqref{eq: loss_bce} weights each data sample equally.
The formulation can be rewritten to factor the freshness of data samples as:
\begin{multline}\label{eq: loss_aoi}
    \mathcal{L}^\prime(\mathcal{N}; \bm{\zeta}) = \sum^{N}_{i=1} - \left[ y_i log (\hat{p}_{y_i}) + (1-y_i) log (1-\hat{p}_{y_i}) \right] e^{-\gamma\Delta_i(t)}. 
\end{multline}
This new loss weights model parameter optimization towards more recent data samples; this function will be denoted as AoI-aware loss formulation.

Let be $\mathcal{N}= \mathcal{N_{\alpha}} \cup \mathcal{N_{\beta}}$, where subscripts denotes recent data and old data. A more general formulation can be written as follows: 
\begin{multline} \label{eq: loss_aoi2}
    \mathcal{L}^{\prime\prime}(\mathcal{N}; \bm{\zeta}) = \alpha\sum^{N_{\alpha}}_{i=1} - \left[ y_i log (\hat{p}_{y_i}) + (1-y_i) log (1-\hat{p}_{y_i}) \right] e^{-\gamma_{\alpha}\Delta_i(t)}  \\
    + \beta \sum^{N_{\beta}}_{j=1} - \left[ y_j log (\hat{p}_{y_j}) + (1-y_j) log (1-\hat{p}_{y_j}) \right] e^{-\gamma_{\beta}\Delta_j(t)}.
\end{multline}
This formulation generalized the previous loss in Eq.~\eqref{eq: loss_aoi}, which is recovered by setting $\alpha=1$, $\beta=0$. For simplicity only the formulation in Eq.~\eqref{eq: loss_aoi} is being considered in this manuscript.

The model's initial deployment and fine-tuning upon drift detection can be described as follows.
Let $\mathcal{N}_{t_0^{\prime}}$ be the initial dataset; $\mathcal{N}_{t_1^{\prime}}$ and $\mathcal{N}_{t_2^{\prime}}$ the datasets collected up to the time $t_{1}^{\prime}$ and $t_{2}^{\prime}$.
At the initial system time $t_0^{\prime}$ the system trains and deploys the model $\hat{\bm{\zeta}}(\mathcal{N}_{t_0^{\prime}})$. At time $t_1^{\prime}>t_0^{\prime}$, the system detects a model drift leading to the fine-tuning of the model.
Because optimization from previous optimal parameters $\hat{\bm{\zeta}}(\cdot)$ leads to faster convergence, we obtain model $\hat{\bm{\zeta}}(\mathcal{N}_{t_1^{\prime}}|\mathcal{N}_{t_0^{\prime}})$ instead of model $\hat{\bm{\zeta}}(\mathcal{N}_{t_0^{\prime}})$.
Upon next drift detection at time $t_2^{\prime}>t_1^{\prime}$ the same thought process can be applied, starting the parameters $\hat{\bm{\zeta}}(\mathcal{N}_{t_1^{\prime}}|\mathcal{N}_{t_0^{\prime}})$ we obtain the parameters $\hat{\bm{\zeta}}(\mathcal{N}_{t_2^{\prime}}|\mathcal{N}_{t_1^{\prime}})$.
When fine-tuning the model, we will use the loss formulation in Eq.~\eqref{eq: loss_aoi}. 

\section{Experimental Setup} \label{sec: experimental setup}

\begin{figure*}[!tbh]
    \centering
    \subfloat
    {\includegraphics[width = 1\linewidth]{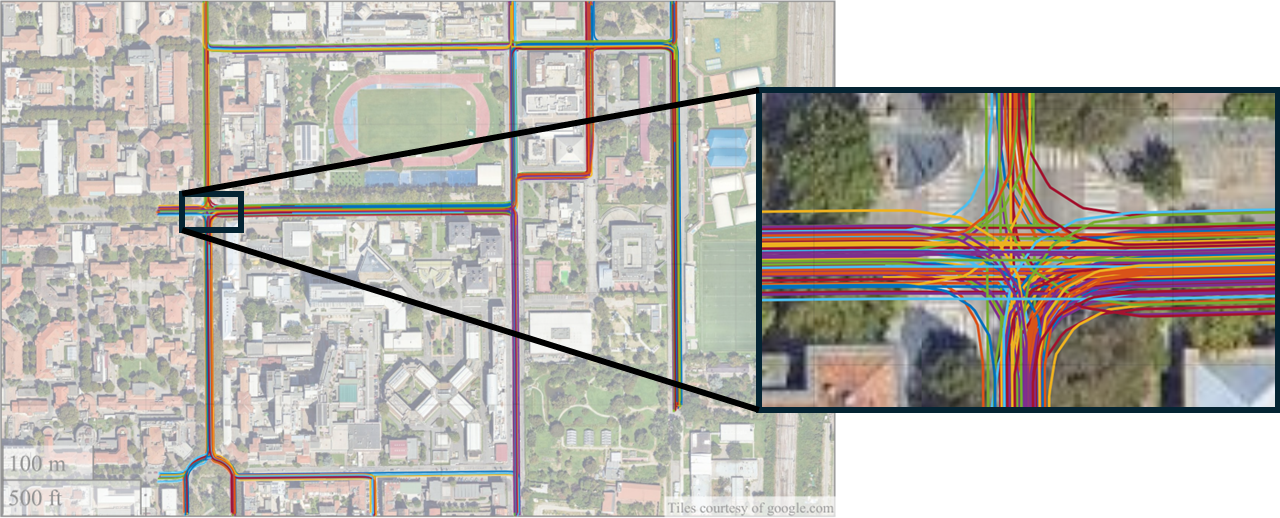}}
    
    \caption{Satellite-view of the modeled scenario, the dashed lines represent the different trajectories of the vehicles. The trajectories are randomly shifted by a small delta in this figure to avoid overlap. Map data \copyright 2024 Google.}
    \label{fig: vehic_satellite}
\end{figure*}

This section describes the experimental setup considered and details the ML training and evaluation procedure.

\subsection{Scenario}\label{subsec: scenario}
The framework is assesed over an area of the city of Milan. The chosen area is delineated by boundary dimensions specified in $670\text{m}\times 550\text{m}$, the boundaries are assumed to absorb EM wave propagation. This region exemplifies a standard 5G/6G microcell and serves as an application scenario for \textit{network digital twinning}. Fig.~\ref{fig: NDT-milan} presents an aerial perspective of the 3D map; the zoom-in shows a snapshot of the simulation with a dynamic wireless environment, where in orange are the parked vehicles in the scene and in blue the moving vehicles.
The ITU recommendation \cite{ITUR2040} is adopted to determine the electromagnetic properties of the object within the scene.

A carrier frequency of $f_c=28$ GHz and a bandwidth of $B=400$ MHz are considered, with the number of subcarriers $N_c$ set to either 512 or 1024. The BS embodies a real-world high-frequency antenna positioned at the top of the main building of the Department of Elettronica, Informazione and Bioingegneria (DEIB) indicated in Fig.~\ref{fig: NDT-milan} (red asterisk). The BS is equipped with a UPA antenna array with $8$ row elements and $16$ column elements $(N_v \times N_h = 128)$, with a spacing of $0.8\lambda$ and $0.5\lambda$ in the vertical and horizontal directions.
The specifications of the antenna elements are delineated according to 3GPP TR 38.901~\cite{3gpp_antenna}. The antenna array is positioned at a height of 21.7 meters, oriented at an angle of 135 degrees relative to the north, and features a downtilt angle of 2 degrees.

\begin{figure} [!bth]
    \centering
    \subfloat{\includegraphics[width = 0.95\linewidth]{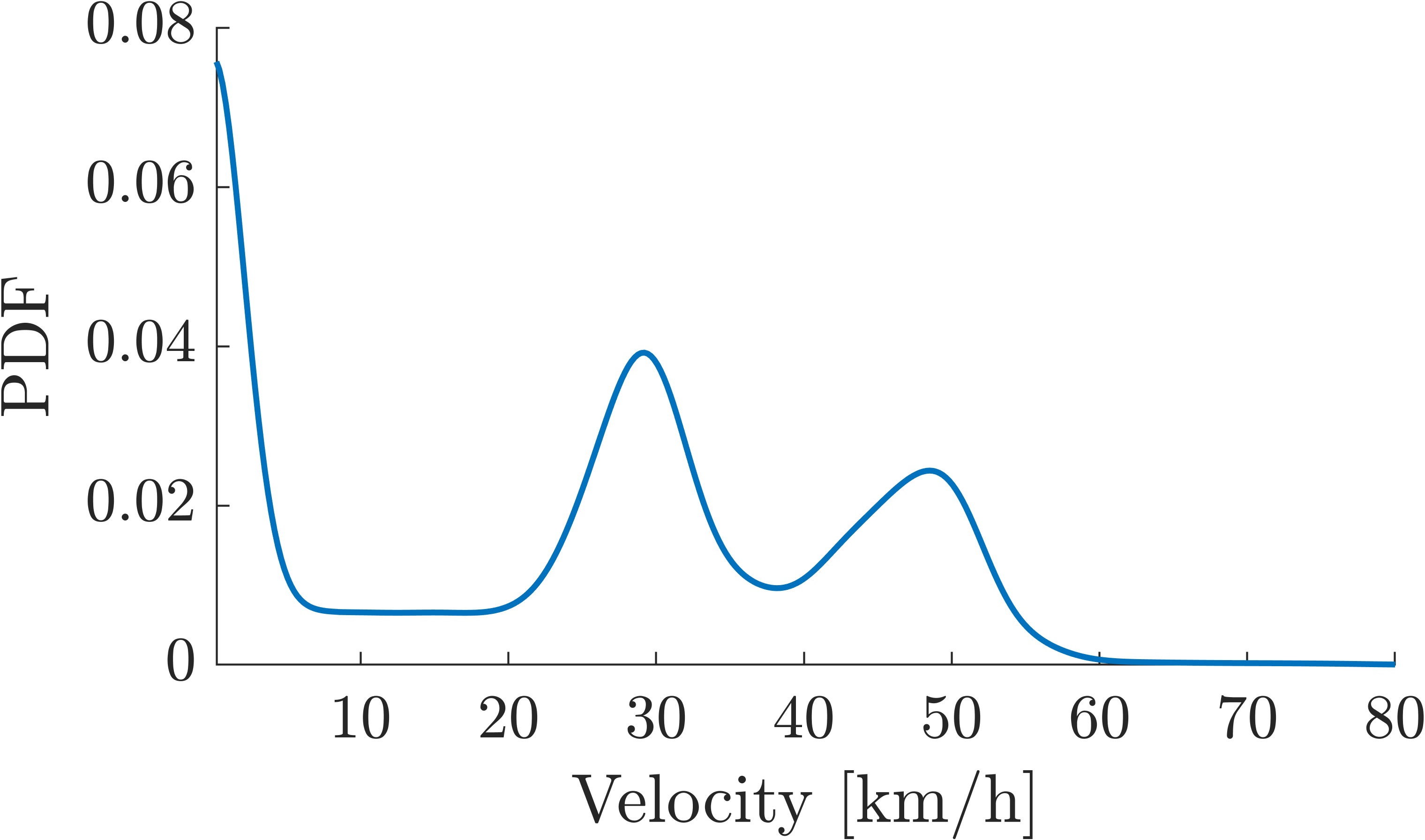}}
    \caption{Empirical probability distribution of the generated vehicular mobility in the analyzed region. The region presents frequent stops, short lanes, and long lanes.}
    \label{fig: speed}
\end{figure}
Based on the specified 3D environment, two scenarios are analyzed:
static scenario, all object are stationary, it includes parked vehicles, shown in Fig.~\ref{fig: NDT-milan}; dynamic scenario, where the mobility has been simulated with SUMO~\cite{SUMO2018}.
The vehicle meshes are obtained from CARLA simulator~\cite{Dosovitskiy17}. The urban scenario presents short and long straight lanes with roundabouts, stoplights, and crossroads. Fig.~\ref{fig: speed} reports the histogram of velocities, which has a high concentration of frequent stops and high speeds (over the speed limit) corresponding to the long lanes. The used mobility model captures a realistic traffic scenario~\cite{vehicular_urban}.

\subsection{Data generation process}\label{subsec: data_generation}
Here, the data generation process using NDT-based raytracing is first detailed, followed by a description of the individual datasets.

\textbf{Raytracing data}. Given the 3D model of the communication environment, raytracing can be used to simulate the propagation paths between each transmitter-receiver antenna pair with raytracing, we use Sionna v0.18~\cite{Hoydis_2023}. Considering the $i$th UE and the $p$th propagation path, its parameters can be written as:
\begin{equation}
    \mathbf{q}_{i, p} = \left[\alpha_{i, p}, \tau_{i, p}, \vartheta_{i, p}, \varphi_{i, p} \right],
\end{equation}
where $\alpha_{i, p} \in \mathbb{C}$ is the complex path coefficient; $\tau_{i,p}, \vartheta_{i, p}, \varphi_{i, p} \in \mathbb{R}$ are the delay, angle of arrival in azimuth and elevation. The raytracing data between the BS and the $i$th UE can be written as: $\mathbf{Q}_i \triangleq [\mathbf{q}_{i, 1}, \mathbf{q}_{i,2}, ..., \mathbf{q}_{i, N_p}]$, where $N_p$ is the number of paths determined by the raytracer. The raytracing data can be obtained from the following link\footnote{\textcolor{red}{GitHub link space [TBD]} }.

\textbf{Grid Dataset} ($\mathcal{G}$). The modeled scenario is divided into non-overlapping $2\times2$ squares forming a grid, each square center corresponds to a potential user position. The generated dataset is denoted as $\mathcal{G}:=\{(\mathbf{Q}_{j}, y_{j}, u_{j}(t_0))\}^{|\mathcal{G}|}_{j=1}$, where $\mathbf{Q}_{j}$ is the raytracing data, $y_{j}$ is the categorical entry associated with the blockage condition, and $u_{j}(t_0)=0$ is the data generation timestamp set to zero. The generation process generates a total of 57000 data samples.
The simulation captures a static wireless environment, i.e., the \textit{fingerprinting} assumption, and is used to initially assess the deployment of ML models.

\textbf{Vehicular Dataset} ($\mathcal{V}$). Synthetic traffic data has been generated with the SUMO simulator~\cite{SUMO2018}. The simulation is run for 300 s, and vehicle positions are sampled every 0.1 s. The simulation contains 4 different types of vehicle: sedan, truck, bus, and hatchback. Each vehicle type has its own 3D meshe and antenna positioned on top. 
In Fig.~\ref{fig: vehic_satellite} we show an example of different trajectories of the vehicles.
For each simulation timestamp the vehicles are moved to their new position within the NDT, and raytracing is performed, generating a total number of $155336$ samples. 
The generated dataset is denoted as $\mathcal{V}:=\{(\mathbf{Q}_m, y_m, u_m(t))\}_{m=1}^{|\mathcal{V}|}$, where $\mathbf{Q}_m$ is the raytracing data, $y_m$ is the categorical entry associated with the blockage condition, and $u_m(t_0)=0$ is the SUMO simulation timestep. The simulation captures a dynamic wireless environment, i.e., Fig.~\ref{fig: speed}, where the movement of the vehicles can significantly affect the propagation paths.

\textbf{Composite Datasets} $(\mathcal{S}_k)$. The following 3 composite datasets are defined as: $\mathcal{S}_k := \{\mathcal{G} \cup \mathcal{V}| t \leq t_k + 10\}$, with $k=1,2,3$ and $t_1=90, t_2=190, t_3=290$. The last 10 seconds will be used to evaluate the models. To further increase the realism of the simulation a random probability to remove the LoS component has been added, modeling the condition in which a non-tracked object obstructs the LoS path.
The goal of this composition is to validate the effectiveness of the AoI-aware loss function in model fine-tuning as described in Sec.~\ref{subsec: aoi-learning}. At time $t_1$ it is assumed a model drift detection and the previous model fine-tuned $\bm{\hat{\zeta}}(\mathcal{G})$, obtaining the model $\bm{\hat{\zeta}}(\mathcal{S}_1 | \bm{\hat{\zeta}}(\mathcal{G}) )$. At time $t_2$, another model drift is assumed to be detected, so fine tuning obtains model $\bm{\hat{\zeta}}(\mathcal{S}_2 | \bm{\hat{\zeta}}(\mathcal{G}) )$. The same process is repeated for $t_3$.

\subsection{Model training and evaluation procedure.} \label{subsec: model train and eval}
\textbf{Statistical Machine Learning.} The i$th$ input sample for SML is defined as $(\bm{\psi}_i, y_i)$, where $\bm{\psi}_i$ is the vector with entries defined in Sec.~\ref{subsec: sml-features} and $y_i$ the blockage label. During training $\bm{\psi}_i$ is directly derived from $\mathbf{Q}_i$, while during testing we synthesize the channel and perform multipath component estimation. We perform a k-fold cross-validation with ten folds~\cite{10.5555/1162264}. A grid search has been performed for hyperparameter tuning. In the following analysis section, we report only the best model. The models are trained in MATLAB (v23.2, R2023b) Update 2.

\begingroup
\renewcommand{\arraystretch}{1.3} 
\begin{table}
\centering
\caption{Table reporting the composite datasets $\mathcal{S}_1, \mathcal{S}_2, \mathcal{S}_3$, each dataset is further split into train, validation, test sets. $I_{trn}, I_{val}, I_{tst}$ refer respectively to the timestamps reserved for training, validation, and testing.}
\label{tab: composite dataset split}
    \begin{tabular}{@{}lllllll@{}}
        \toprule 
            & $I_{trn}$ & $I_{val}$ & $I_{tst}$ \\ 
        \midrule 
        
        $\mathcal{S}_1$ & $[0, 80)$ & $[80, 90)$ & $[90, 100)$ \\ 
        \addlinespace[2 pt]
        $\mathcal{S}_2$ & $[0, 180)$ &  $[180, 190)$ &  $[190, 200)$  \\ 
        \addlinespace[2 pt]
        $\mathcal{S}_3$ & $[0, 280)$ & $[280, 290)$ & $[290, 300)$ \\ 
        \midrule
        \bottomrule
    \end{tabular}
\end{table}
\endgroup

\textbf{Deep Learning.} The i$th$ input sample for DL is defined as $(\mathbf{P}_i, y_i, u_i(t))$ in Sec.~\ref{subsec: adcpm-def}, where $\mathbf{P}_i$ is the ADCPM defined in Eq.~\eqref{eq:adcpm}, $y_i$ the blockage label and $u_i(t)$ the data generation timestamp. We apply hold-out evaluation with validation splits, i.e., the dataset is split into train, validation, and test sets. For datasets $\mathcal{G}, \mathcal{V}$, the split is performed with random sampling with the following ratio $[0.7, 0.2, 0.1]$. The splits of the composite datasets are reported in Table~\ref{tab: composite dataset split}. To avoid model overfitting, we apply the early stopping technique with patience of 5 epochs \cite{bai2021understandingimprovingearlystopping}. The models can be trained up to a maximum of $N_e=200$ epochs, and early stopping terminates the training approximately in the interval of $(10, 20)$ epochs. The model is trained with batch size $N_b=32$, learning rate $lr = 10^{-4}$, Adam optimizer with momentum $\beta_1=0.9$, $\beta_2=0.999$ \cite{kingma2017adammethodstochasticoptimization}. We choose to use the architecture of the ResNet34 model \cite{He_2016}, a popular Convolutional Neural Network (CNN) architecture, the fully connected classifier head used has the following parameters: $1$st layer 512 neurons, $2$nd layer 256 neuros, and lastly the output probabilities are obtained with a sigmoid activation function \cite{10.5555/1162264}. 
Given model output logits $\nu$, i.e., outputs from the model's last layer, the sigmoid function:
\begin{equation} \label{eq: sigmoid}
    \sigma(\nu) = \frac{1}{1+e^{-\nu}}.
\end{equation}
The model is implemented in Python (v3.10.11), Pytorch (v2.2.0), and single floating point precision. We note that while we evaluate our approach for the specific ResNet34 architecture, it can be generalized to almost any other architecture.

\section{Numerical Analysis} \label{sec: num analysis} 
\begin{figure} [t]
    \centering
    \includegraphics[width=0.9\columnwidth]{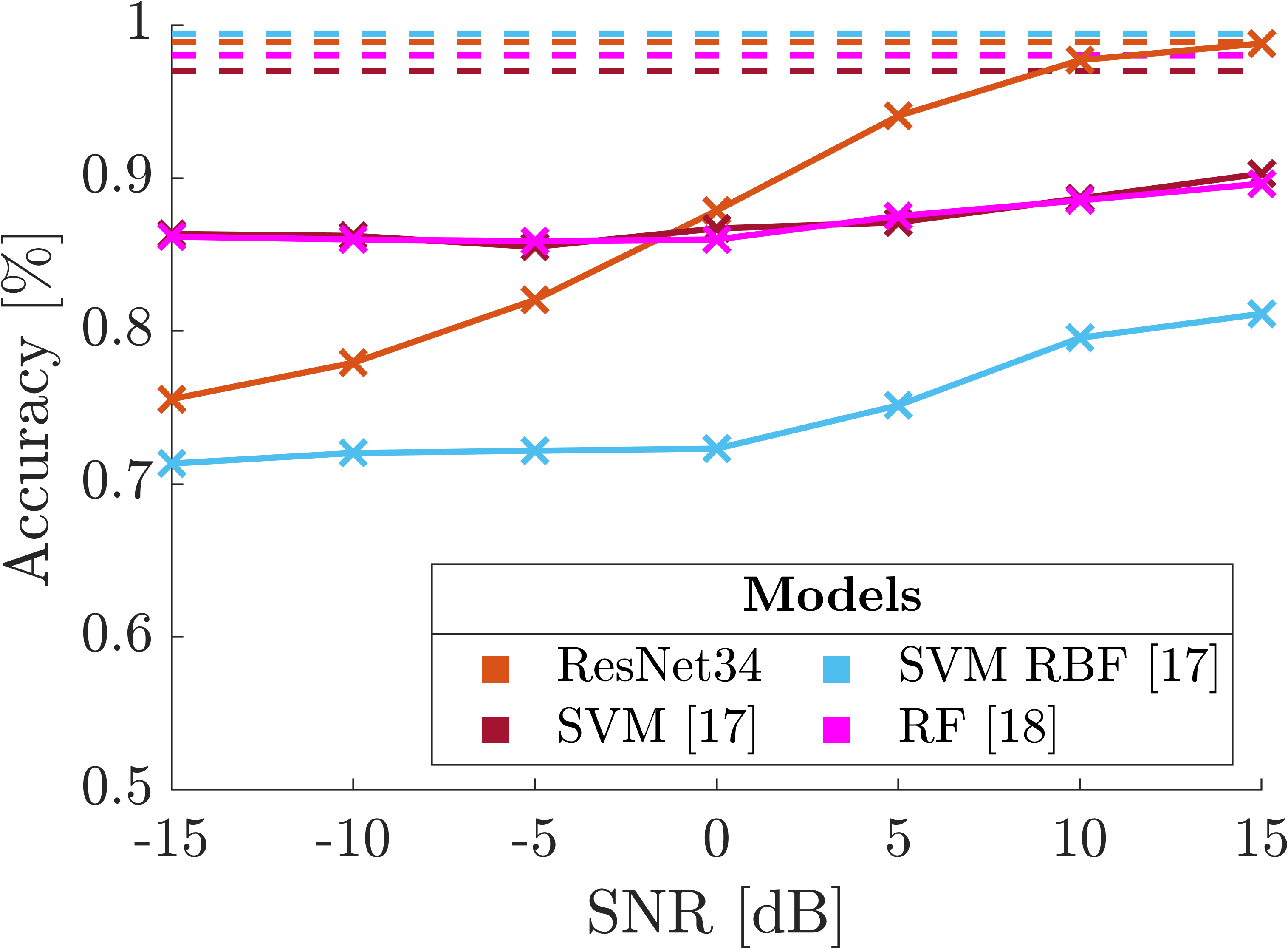}
    \caption{ML model performance trained with grid dataset $\mathcal{G}$, the evaluation is carrier out over the test split. Dashed lines are the noise-free bound. Solid lines are the noisy data with power control and high-resolution parameter estimation.}
    \label{fig: accuracy_ml_dl}
\end{figure}
Considering the grid-based data $\mathcal{G}$ and the MIMO-OFDM configuration with $N_c=512$ subcarriers. The SML algorithms presented in~\cite{8968748, 10359439} are evaluated, with path parameters obtained through the MUSIC algorithm~\cite{doi:https://doi.org/10.1002/9781119294016.ch7}. The test set evaluation results are presented in Fig.~\ref{fig: accuracy_ml_dl}.
The dashed lines represent the theoretical upper bounds, corresponding to an ideal noiseless scenario and complete knowledge of multipath components; the accuracy of any algorithm exceeds $98\%$.
In a more realistic condition, with noise and multipath estimation error, there is a big discrepancy between the theoretical bound and the model performance.
The DL model is robust in terms of the difference between the NDT synthetic channel and the one with noise; it shows a downward trend corresponding to the decrease in SNR. DL performance falls below SML models at $\text{SNR} = 0 \, \text{dB}$.
While SML learns to separate the distribution in the training samples, the DL models first learn a latent representation of the channel measurements and then classify them.

\begin{figure} [tbh]
    \centering
    \includegraphics[width=0.9\columnwidth]{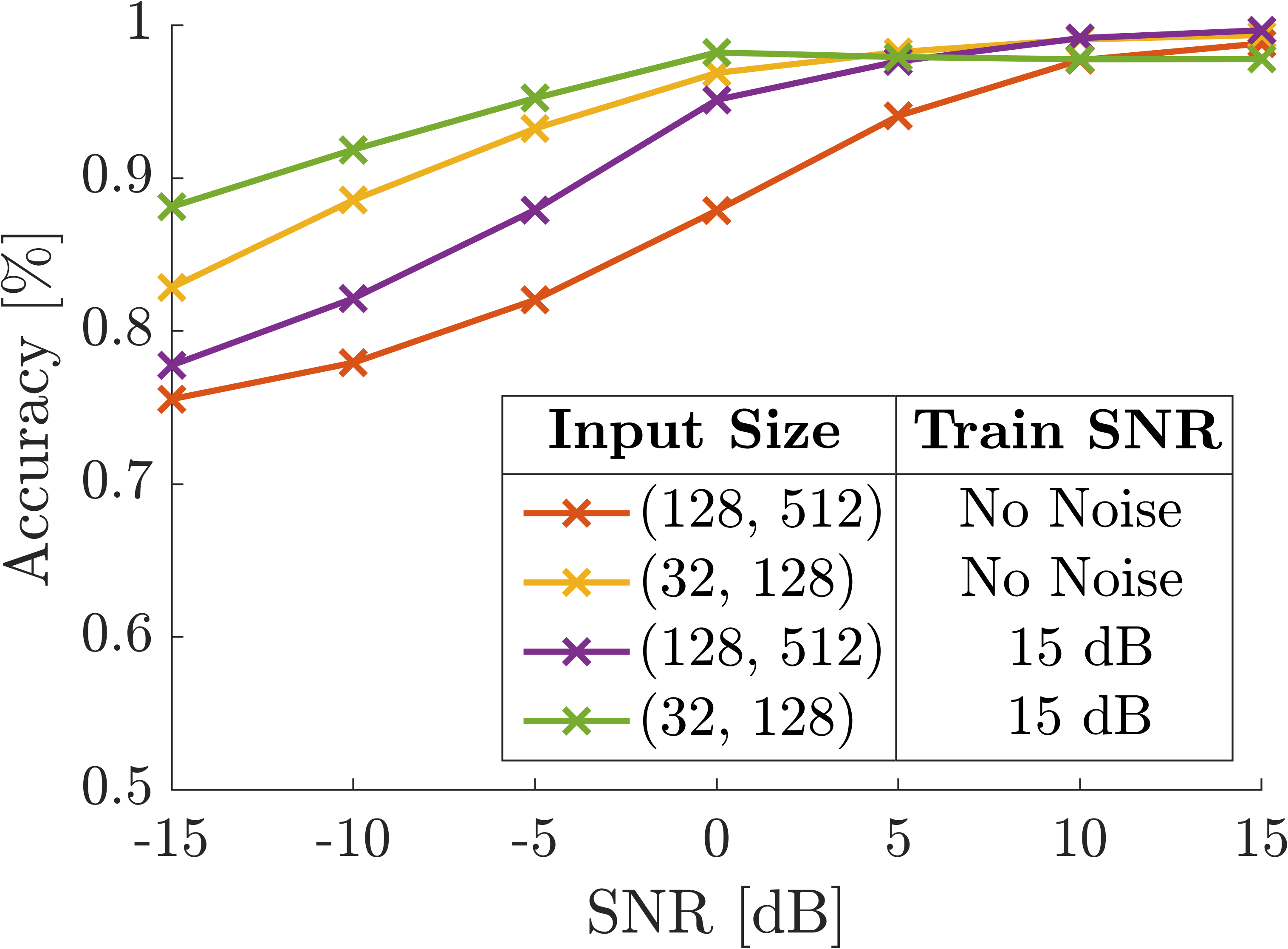}
    \caption{We conduct an ablation study to evaluate the efficacy of data augmentation and resolution reduction. The models are trained and evaluated with the grid dataset $\mathcal{G}$. The initial dimension of the ADCPM is $(128, 512)$, the reduced resolution is $(32, 128)$. Data augmentation is performed by simulating power control with a $\text{SNR} = 15 \text{ dB}$.}
    \label{fig: ablation}
\end{figure}

We determine two strategies to improve DL model performance:
\begin{itemize}
    \item[i)] \textit{resolution reduction};
    \item[ii)] \textit{data augmentation}.
\end{itemize}
The proposed resolution reduction \textit{(i)} for the angle delay power profile is described in detail in~\ref{appendix: cnn cost}; data augmentation \textit{(ii)} is a technique used to enhance pre-existing data samples to create new data samples that can improve model optimization and generalization, we perform it by simulating power control with $\text{SNR}=15 \text{ dB}$ in the training samples. 

\noindent To validate the proposed techniques, an ablation study is performed.
In Fig.~\ref {fig: ablation}, the red line is the ResNet34 presented (our baseline) in the previous figure. The purple line is the variant with data augmentation during the training, the yellow line is the model that has its input resolution reduced from $(128, 512)$ to $(32, 128)$ with a $(4, 4)$ max-pooling. The combination of both techniques results in the green line, showing 10\% improvement when SNR is below 0 and 5\% improvement when SNR is 5 dB compared to the baseline (red).

The reduction in resolution not only improves model accuracy, but it also leads to improved computational cost of the model. The latency speedup factor can be calculated as follows:
\begin{equation}
    S = \frac{C_{old}}{C_{new}},
\end{equation}
where $C_{old}$ is the old computational cost and $C_{new}$ is the computational cost after resolution reduction. The computational cost is measured in Floating Point Operations Per Second (FLOPs). It should be noted that DL models are trained and designed to operate on a fixed input size, while 5G NR introduces flexible numerology~\cite{8911694}.
The max-pooling can be used to dynamically adapt the channel measurements to a fixed size. This reduction is fixed to $(32, 128)$ and the number of subcarriers is increase to $N_c = 1024$ in the next experiment.
In Table~\ref{tab: cost reduction} the speedup factor for $N_c = [512, 1024]$ is reported, when the input reduction is reduced to $(32, 128)$ from $(128, 512)$ the inference is $16$ times faster than with its original dimensions, when the reduction is from $(128, 1024)$ the inference speed is $32$ times faster.

\begingroup
\renewcommand{\arraystretch}{1.3}
\begin{table}
    \centering
    \caption{Deep Learning model speedup factor.}
    \begin{tabular}{ccc}
        \toprule
          $\mathbf{H}$ Dimension & Reduced Input Size & Speedup Factor\\
        \midrule
            $(128, 512)$ & $(32, 128)$ & 16 \\ 
            $(128, 1024)$ & $(32, 128)$ & 32 \\ 
        \midrule

        \bottomrule
    \end{tabular}

    \label{tab: cost reduction}
\end{table}
\endgroup

\begin{figure*} [tbh]
    \centering
    \subfloat[Output distribution at SNR 15 dB]{\includegraphics[width = 0.33\linewidth]{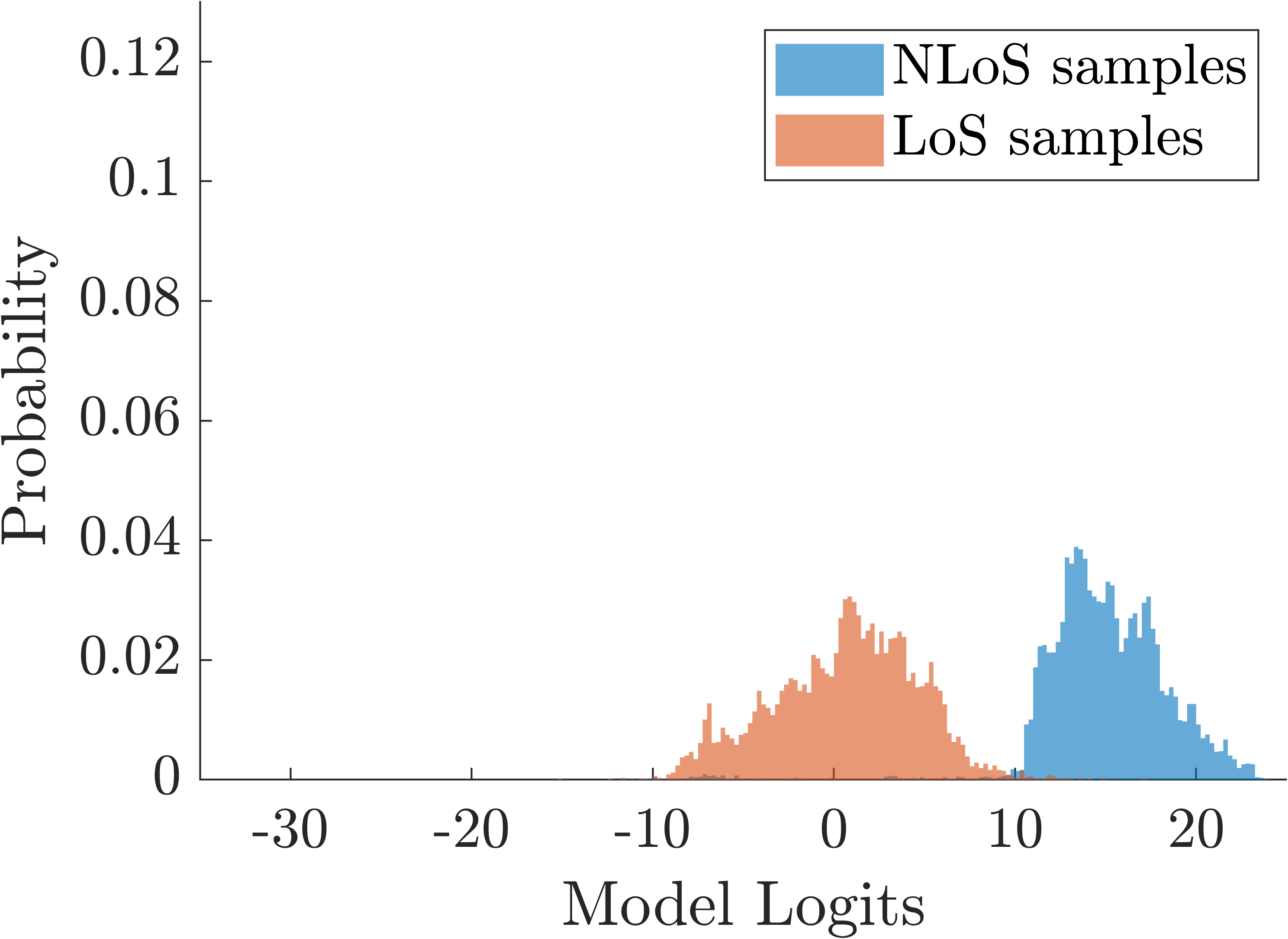}\label{subfig: logits+15dB}}
    \subfloat[Output distribution at SNR 0 dB]{\includegraphics[width = 0.33\linewidth]{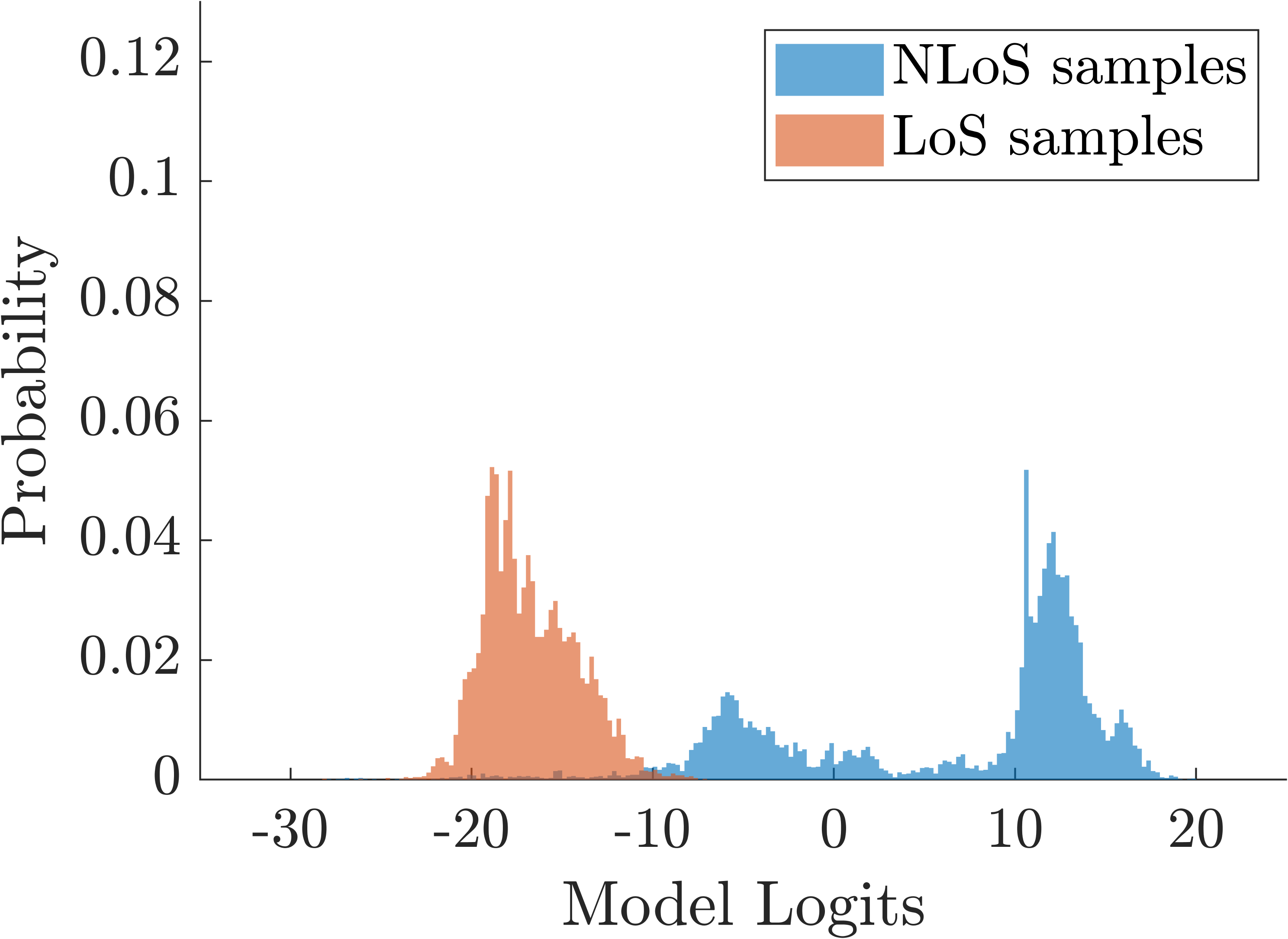}\label{subfig: logits+0dB}}
    \subfloat[Output distribution at SNR -15 dB]{\includegraphics[width = 0.33\linewidth]{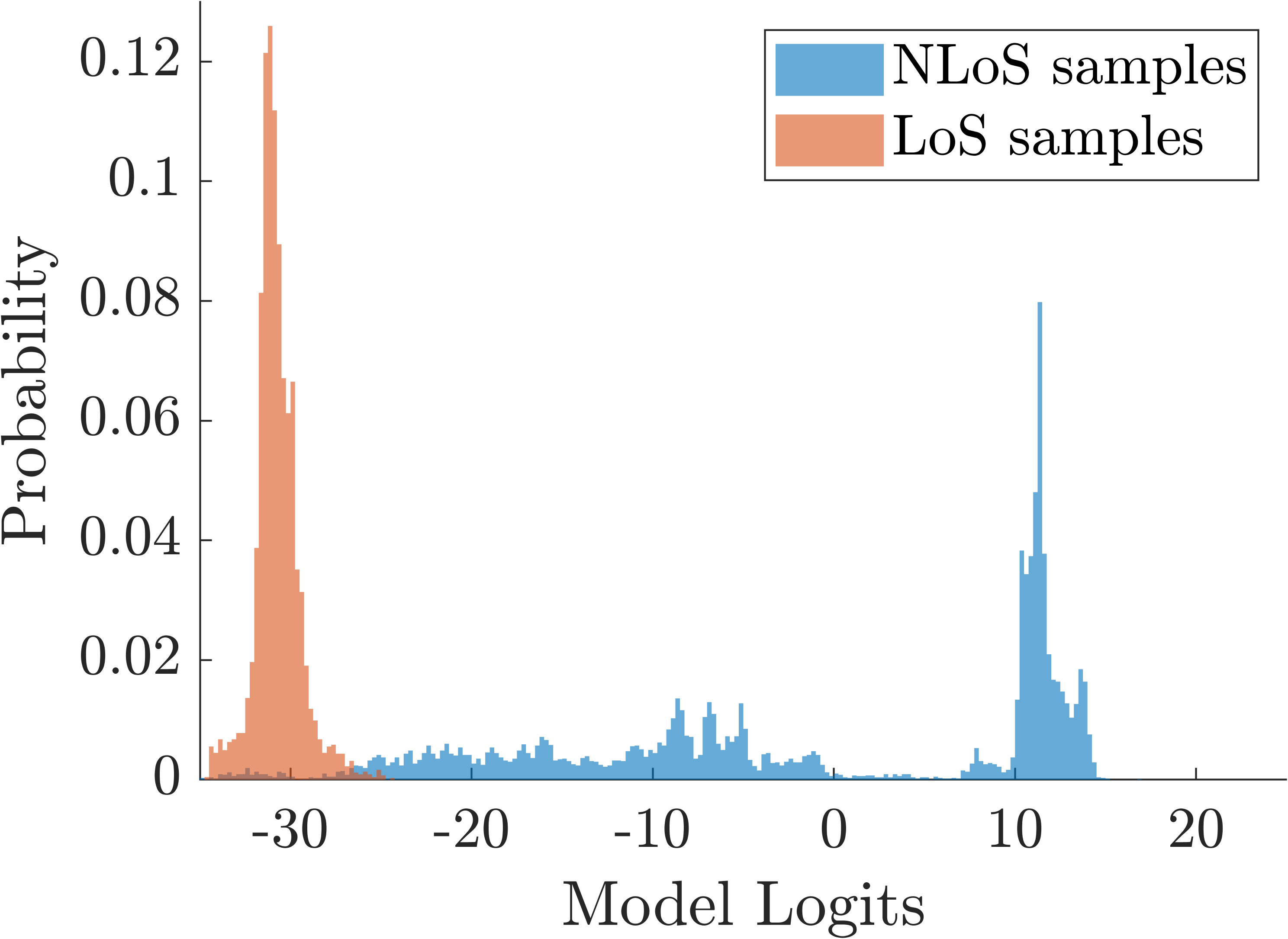}\label{subfig: logits-15dB}}
    \caption{Model logits output over test subset of vehicular dataset before softmax function. In this experiment the number of wireless subcarriers is set to 1024.}
    \label{fig: logits}
\end{figure*}

The previously trained model (green curve in Fig.~\ref{fig: ablation}) is evaluated on the vehicular dataset $\mathcal{V}$ with 1024 subcarriers.
The model fails to accurately identify the blockage condition. 
The output logit distribution $\nu$ (raw, non-normalized predictions, before classification threshold) in Fig.~\ref{fig: logits} shows a distribution shift.
One can dynamically analyze the ROC curve~\cite{FAWCETT2006861} to find the optimal classification threshold in conditions such as the one in Fig.~\ref{subfig: logits+15dB}, for the other two cases the application of the sigmoid function in Eq.~\eqref{eq: sigmoid} creates numerically unstable values leading to an unfeasible computation of the ROC curve. Therefore, model fine-tuning is necessary.

In Sec.~\ref{subsec: DL model}, AoI metrics have been introduced as a tool to fine-tune DL models.
The age violation threshold $\Gamma$ is set to $0.005$ and the decay rates $\gamma_1 = 0.4, \gamma_2 = 0.2, \gamma_3 = 0.1, \gamma_4 = 0.01, \gamma_5 = 0.05$ are evaluated.
At time $t_1=90$, the experiments assume the occurrence of a model drift, initiating fine-tuning using the model parameters $\hat{\bm{\zeta}}(\mathcal{G})$ and dataset $\mathcal{S}_1$. Subsequently, at time $t_2=190$, another instance of model drift is detected, and the previously fine-tuned model is further adapated using dataset $\mathcal{S_2}$, yielding $\hat{\bm{\zeta}}(\mathcal{S}_2|\hat{\bm{\zeta}}(\mathcal{S}_1))$. At time $t_3=290$, a third drift is observed, prompting an additional fine-tuning step that produces $\hat{\bm{\zeta}}(\mathcal{S}_3|\hat{\bm{\zeta}}(\mathcal{S}_2))$.
Table~\ref{tab: AOI-performance} reports the model accuracy at different decay rates and details the number of samples used in training, validation, and testing. The experiment is carried out with $N_c = 1024$ and evaluation with $\text{SNR}=-15 \text{ dB}$. This fine-tuning strategy obtains DL models with a performance that exceeds $98\%$ in accuracy.

\begingroup
\renewcommand{\arraystretch}{1.3}
\begin{table} [tbh]
    \centering
    \caption{Model AoI-aware fine-tuning performance. $N_{trn}, N_{val}, N_{tst}$ refers to the number of train, validation, and test samples. The drop in the number of samples for $\mathcal{S}_1$ is caused by the inclusion and exclusion of the grid dataset $\mathcal{G}$.}
    \begin{tabular}{llcccc}
        \toprule
         &  & Accuracy \% & $N_{trn}$ & $N_{val}$ & $N_{tst}$\\
        \midrule
            & $\mathcal{S}_1$ & 98.55 & 90193 & 4245 & 4357 \\ 
         $\gamma_1=0.01$ & $\mathcal{S}_2$ & 96.82 & 137761 & 6312 & 6739 \\ 
            & $\mathcal{S}_3$ & 99.35 & 205049 & 7288 & 7808\\ 
        \midrule
            & $\mathcal{S}_1$ & 98.05 & 90193 & 4245 & 4357 \\ 
         $\gamma_2=0.05$ & $\mathcal{S}_2$ & 96.57 & 45815 & 6312 & 6739 \\ 
            & $\mathcal{S}_3$ & 99.00 & 64754 & 7288 & 7808\\ 
        \midrule
            & $\mathcal{S}_1$ & 98.46 & 17703 & 4245 & 4357 \\ 
         $\gamma_3=0.1$ & $\mathcal{S}_2$ & 96.67 & 22785 & 6312 & 6739 \\ 
            & $\mathcal{S}_3$ & 98.24 & 29146 & 7288 & 7808\\ 
        \midrule
            & $\mathcal{S}_1$ & 98.46 & 6785 & 4245 & 4357\\ 
         $\gamma_4=0.2$ & $\mathcal{S}_2$ & 94.85 & 9315 & 6312 & 6739 \\ 
            & $\mathcal{S}_3$ & 98.29 & 10938 & 7288 & 7808\\ 
        \midrule
            & $\mathcal{S}_1$ & 98.25 & 1267 & 2187 & 2170\\ 
         $\gamma_5=0.4$ & $\mathcal{S}_2$ & 95.01 & 1870 & 6312 & 6739\\ 
            & $\mathcal{S}_3$ & 98.48 & 2132 & 7288 & 7808\\ 
        \midrule
        \bottomrule
    \end{tabular}

    \label{tab: AOI-performance}
\end{table}
\endgroup

The integration of AoI-loss and thresholding reduces the number of training samples, resulting in only a marginal degradation in performance. Specifically, a comparison between configurations $\gamma_1$ and $\gamma_5$ shows an accuracy reduction of approximately $1\%$, which is negligible given that $\gamma_5$ utilizes only $1\%$ of the total training data from $\mathcal{S}_3$.

\section{Discussion and future directions} \label{sec: discussion}
This section aims to summarize the main observations from our simulations and provide future research directions.

Model-based solutions, such as those presented in~\cite{9253591, 4510769}, typically rely on simplified assumptions regarding the environment, signal propagation characteristics, or user behavior. These assumptions fail to capture the complexity and variability of dynamic deployment scenarios. While data-driven method often obtains more accurate results, they are limited by the availability and the variety of the data collected during measurement campaigns~\cite{8968748, 10121016, SINGH2023102118}. A critical issue in related data-driven works is the implicit assumption that the learned models will remain valid over time. As demonstrated by Vela et al.~\cite{Vela2022}, dynamic environments often lead to distribution shifts, which can significantly degrade model performance.
Integrating a \textit{digital twin} with data-driven frameworks enables a more flexible and adaptive learning approach. By continuously ingesting real-time data and updating the models, the system mitigates the limitations of traditional measurement campaigns. It increases data variety, offering improved accuracy and robustness in dynamic environments.

The implementation of the proposed NDT requires the integration of a digital map and a raytracing simulator. The construction of a 3D digital map is relatively tractable with existing tools or is readily available~\cite{OpenStreetMap}. Emerging trends in areas such as autonomous driving and the Internet of Things (IoT) are expected to further enhance the availability and fidelity of such maps~\cite{ZHAO2023101974}. Extensive research in wireless propagation modeling using raytracing techniques has led to the development of increasingly accessible and computationally efficient software~\cite{Hoydis_2023, zhu2024realtime}. Recent advancements in the semiconductor industry have remarkably improved the computational capabilities of low-power devices, resulting in cost-effective deployments of AI models. Table~\ref{tab: nvidia_modules} reports the inference time for processing 32 samples with NVIDIA Jetson modules, demonstrating the proposed approach's suitability for the application. Further performance and efficiency gains may be achieved through model compression techniques~\cite{9643539} or switching in model backbones beyond the one proposed~\cite{Sandler_2018_CVPR}. Once the NDT infrastructure is established, it can be readily scaled across multiple deployment sites, yielding substantial reductions in both capital expenditures (CAPEX) and operational expenditures (OPEX) for mobile network operators, as demonstrated in~\cite{Cui2025-am}.

\begin{table}[btp]
    \centering
    \caption{Floating-point operations per second (FLOPS), cost, and inference time of selected NVIDIA Jetson modules. Inference time is measured using the ResNet34 architecture with max-pooling to $(32, 128)$ and 32 samples (batch-size).}
    \label{tab: nvidia_modules}
    \resizebox{\linewidth}{!}{
    \begin{tabular}{lccc}
        \toprule
        \textbf{GPU Module} & \makecell{\textbf{Throughput}\\(FP32)} & \makecell{\textbf{Cost}\\(USD)} & \makecell{\textbf{Inference}\\\textbf{Time} (ms)} \\
        \midrule
        Jetson Nano & 235.8 GFLOPS & \$99 & 6.36 \\
        Jetson Orin Nano 4GB & 640 GFLOPS & \$199 & 2.34 \\
        Jetson AGX Orin 32GB & 3.33 TFLOPS & \$899 & 0.45 \\
        \bottomrule
    \end{tabular}
    }
\end{table}

The proposed method is designed for urban micro-cell-level deployments, in line with emerging trends in 5G/6G network architecture, where micro-cell densification is favored. The deployment of multiple localized models often incurs lower cost and greater flexibility than attempting to generalize across a broad, heterogeneous environment~\cite{7422408}. Scaling up the approach to larger geographical areas would lead to exponential increases in raytracing simulation complexity and data processing overhead. From a practical standpoint, deploying multiple (scaling-out) site-specific models is more computationally efficient and cost-effective than constructing a single large-scale model.

This integration of AoI with NDTs and ML models aligns with the class of \textit{continual learning strategies}~\cite{10444954}, specifically those based on regularization. The AoI loss function in~\eqref{eq: loss_aoi} introduces a temporal weighting mechanism that prioritizes recent samples during model updates. During the backward pass, more recent observations are assigned higher importance, allowing the model to adapt more rapidly to current conditions while gradually discounting outdated information.
Temporal context of older data samples can be extended through hybridization with \textit{replay-based methods}~\cite{NIPS2017_f8752278, DBLP:journals/corr/abs-1902-10486}, as proposed in Eq.~\eqref{eq: loss_aoi2}. Future research should prioritize identifying optimal hyperparameters within Eq.~\eqref{eq: loss_aoi2}, particularly those governing the balance between historical and newly acquired data.

The authors acknowledge that continual learning introduces the risk of \textit{catastrophic forgetting}~\cite{Chen2018}---a phenomenon where the model loses previously acquired knowledge when updated with new data~\cite{Chen2018}. The NDT system enables continuous model monitoring and validation. In the event of unexpected behavior, the system can automatically revert to previously validated models or deploy traditional methods as a fallback mechanism.
The integration of \textit{replay-based methods} in Eq.~\eqref{eq: loss_aoi2} enables the system to maintain exposure to earlier data distributions during training, potentially preventing \textit{catastrophic forgetting}. Further investigation is necessary to systematically assess the long-term stability and resilience of the proposed strategy.
\section{Conclusion} \label{sec: conclusion}
This study presents an NDT-based data collection, providing a scale-out approach for site-specific \textit{data-driven} models. It evaluates training with NDT-generated synthetic data and the corresponding performance, highlighting the impact of an accurate evaluation.
In addition, resolution reduction has been proposed to handle 5G NR flexible numerology and inference with a speedup of up to 32 times. 
The AoI metric is integrated into the collected data samples, and an AoI-aware loss function is proposed to fine-tune deep learning models. The simulation shows that with only $1\%$ of the available data, we can still achieve over $98\%$ accuracy.
In the future, the proposed NDT system will facilitate closed-loop monitoring of machine learning model performance, enabling proactive detection and mitigation of performance degradation. This capability supports the development of robust and reliable systems that can sustain consistent performance throughout their operational lifecycle.

\section*{Declaration of competing interest}
The authors declare that they have no known competing financial interests or personal relationships that could have appeared to influence the work reported in this manuscript.

\section*{Acknowledgment}
The authors acknowledge the help of the Laboratorio di Simulazione Urbana ``Fausto Curti'' of the Department of Architecture and Urban Studies of Politecnico di Milano in the provision of the highly detailed urban digital twin maps used in this work.

The work of Francesco Linsalata, Silvia Mura, and Umberto Spagnolini was partially supported by the European Union under the Italian National Recovery and Resilience Plan (NRRP) of NextGenerationEU, partnership on “Telecommunications of the Future” (PE00000001 - program “RESTART”, Structural Project 6GWINET).

\appendix

\section{Convolutional Layers Cost}\label{appendix: cnn cost}
\begin{figure} [!t]
    \centering
    \subfloat[Sample in LoS condition.]{\includegraphics[width=0.5\linewidth]{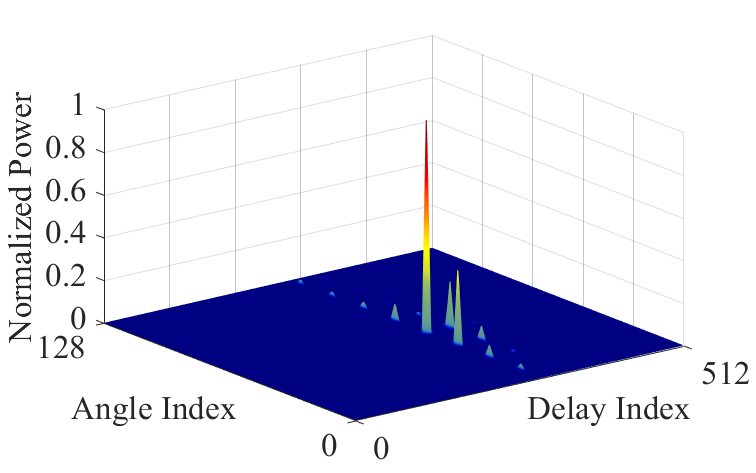}\label{subfig: adcpm_los}}
    \subfloat[LoS sample after $(4, 4)$ max-pooling]{\includegraphics[width=0.5\linewidth]{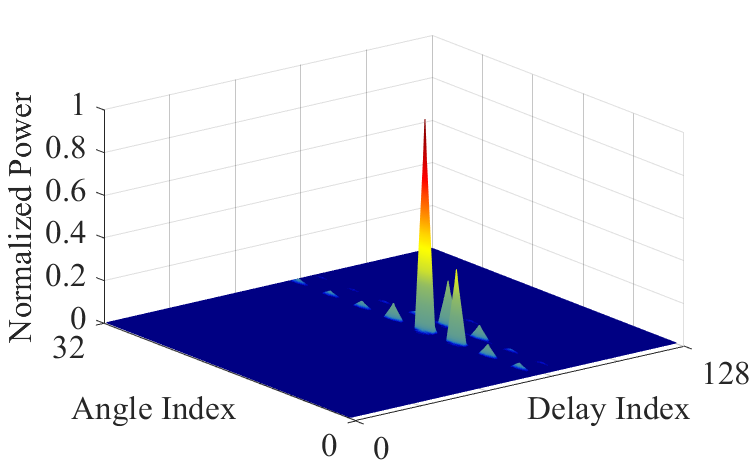}\label{subfig: adcpm_downsampled}}
    \caption{\protect\subref{subfig: adcpm_los} Angle delay power profile of a wireless channel sample in LoS condition with normalized power. \protect\subref{subfig: adcpm_downsampled} The same sample after resolution reduction with a max-pooling with kernel $(4,4)$.}
    \label{fig: ADCPM-sample}
\end{figure} 

The computational cost of DL models scales with channel measurement resolution. Large antenna arrays and flexible numerology in 5G systems significantly increase the resolution, limiting the feasibility of DL models in real-time applications~\cite{CC-complexity}. 
For massive MIMO-OFDM systems with UPA in the BS, the angle delay power profile (ADCPM) is concentrated on specific positions in the horizontal angle-delay domain or vertical angle-delay domain, leading to sparse matrices; detailed derivation is found in~\cite{9364875}. In Fig.~\ref{subfig: adcpm_los}, we show ADCPM samples in LoS condition; the sparsity of the representation leads to unnecessary computation overhead in the feature extraction layers of CNNs for blockage detection. Thus, we propose the max-pooling operation to reduce the resolution of ADCPM samples. Fig.~\ref{subfig: adcpm_downsampled} shows the previous ADCPM samples in LoS after a max-pooling with kernel $(4,4)$. It can be observed that the resolution reduction preserves the characteristics of the original sample. 
The following is a derivation of the computation cost of CNNs. 

CNNs are composed of a series of convolutional layers for feature extraction and a classifier head, where the classifier head generally accounts for $5-10\%$ of the total cost of the network~\cite{7299173}. If we consider a CNN architecture without special layer connections, it is possible to write the total computational complexity of all convolutional layers as:
\begin{equation}
    O\left(\sum^D_{q=1} n_{q-1} d_q^W d_q^H n_q m_q^W m_q^H\right),
\end{equation}
where l is the index of the convolutional layer, d is the depth (number of convolutional layers), $n_{q-1}$ is the number of input channels and $n_q$ is the number of convolutional filters in the $q$th layer, $d_q^W$, $d_q^H$ are the spatial dimensions of the convolutional filters and $m_q^W$, $m_q^H$ are the spatial dimensions of the output feature maps. The superscript $W, H$ indicates the width and height, respectively. The output dimensions of the feature map at the $q$th layer can be written as:
\begin{equation}
    m_q^W = q\left(d_q^W, m_{q-1}^W\right),
\end{equation}
\begin{equation}
    m_q^H = q \left(d_q^H, m_{q-1}^H \right),
\end{equation}
where $q(\cdot)$ is the function that determines the new dimension after the convolutions. 
Consider an ADCPM as input on the first convolutional layer with dimensions $(N_v N_h, N_c)$, the initial number of channels is $n_0 = 1$, and the input spatial dimension is $m_0^W = N_v, m_0^H = N_h$. If we reduce the initial spatial dimensions by constant factors $(a, b)$, we can derive that the final computational cost will be $\frac{1}{ab}$ of the original cost.

\bibliographystyle{IEEEtran}
\bibliography{bibliography} 

\vskip +1\baselineskip plus +1fil
\begin{wrapfigure}{l}{25mm} 
    \includegraphics[width=1in,height=1.25in,clip,keepaspectratio]{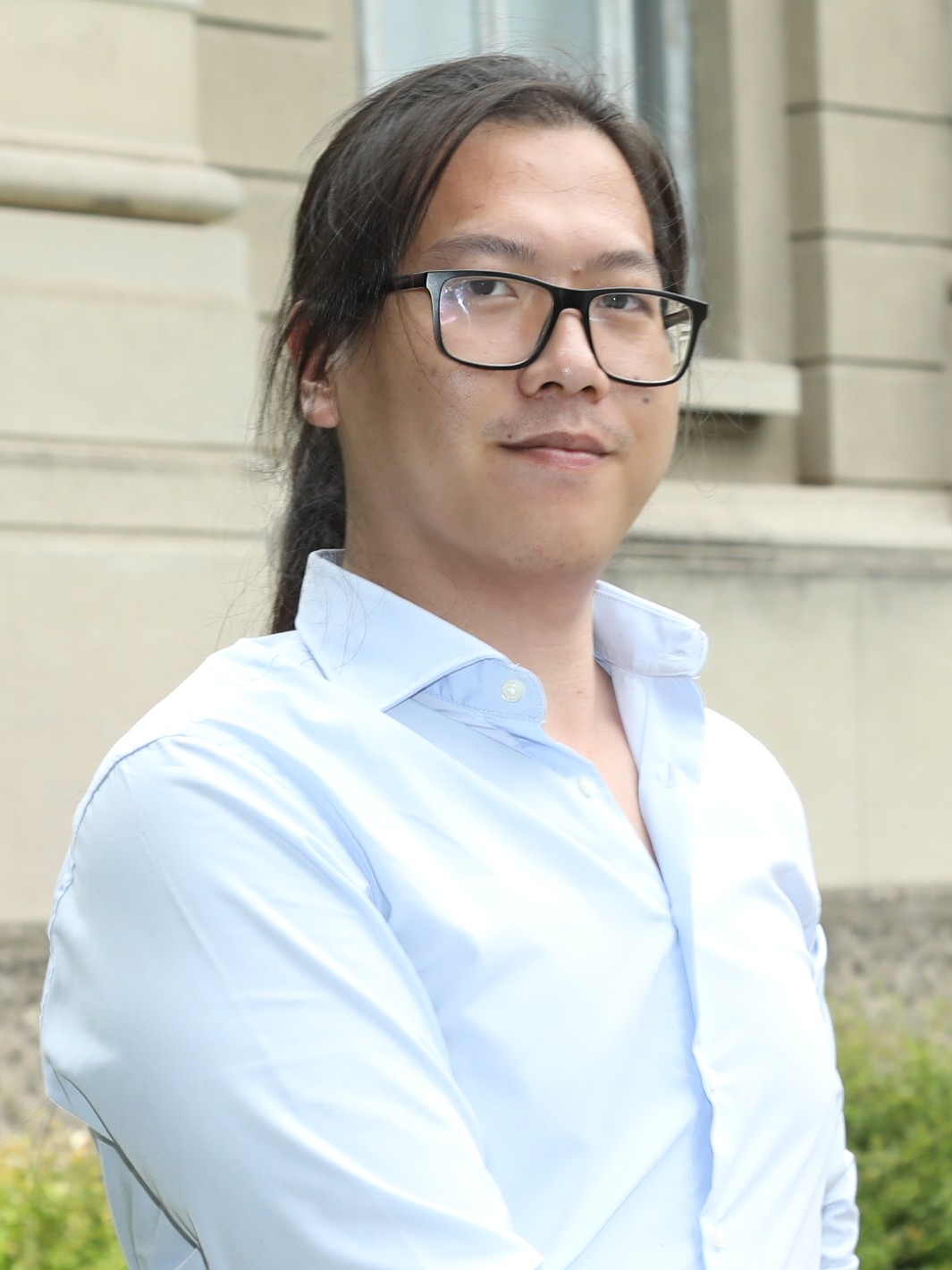}
  \end{wrapfigure}\par
  \noindent \textbf{Michele Zhu} received the B.Sc. in information engineering from Università di Padova, Padua, Italy, and the M.Sc. degree in computer science and engineering from Politecnico di Milano, Milan, Italy. He started pursuing a Ph.D. degree in information technology in September 2023. He is involved in research projects in the Joint Laboratory between Huawei Technologies Italia, Milan, and Politecnico di Milano. \par
  
\vskip +1\baselineskip plus +1fil
\begin{wrapfigure}{l}{25mm} 
    \includegraphics[width=2.5cm,height=3.33cm,clip,keepaspectratio]{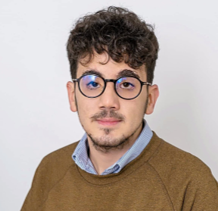}
  \end{wrapfigure}\par
  \noindent \textbf{Francesco Linsalata} received M.Sc and PhD degrees cum laude in Telecommunication engineering from Politecnico di Milano, Milan, Italy, in 2019 and 2022, respectively. He is a researcher at the Dipartimento di Elettronica, Informazione e Bioingegneria, Politecnico di Milano. His main research interests focus on V2X and UAV communications, Integrated Communication and Sensing, and physical layer design for 6G wireless networks. He was co-recipient of one best-paper award and recipient of one best student paper award. \par

\vskip +1\baselineskip plus +1fil
\begin{wrapfigure}{l}{25mm} 
    \includegraphics[width=2.5cm,height=3.33cm,clip,keepaspectratio]{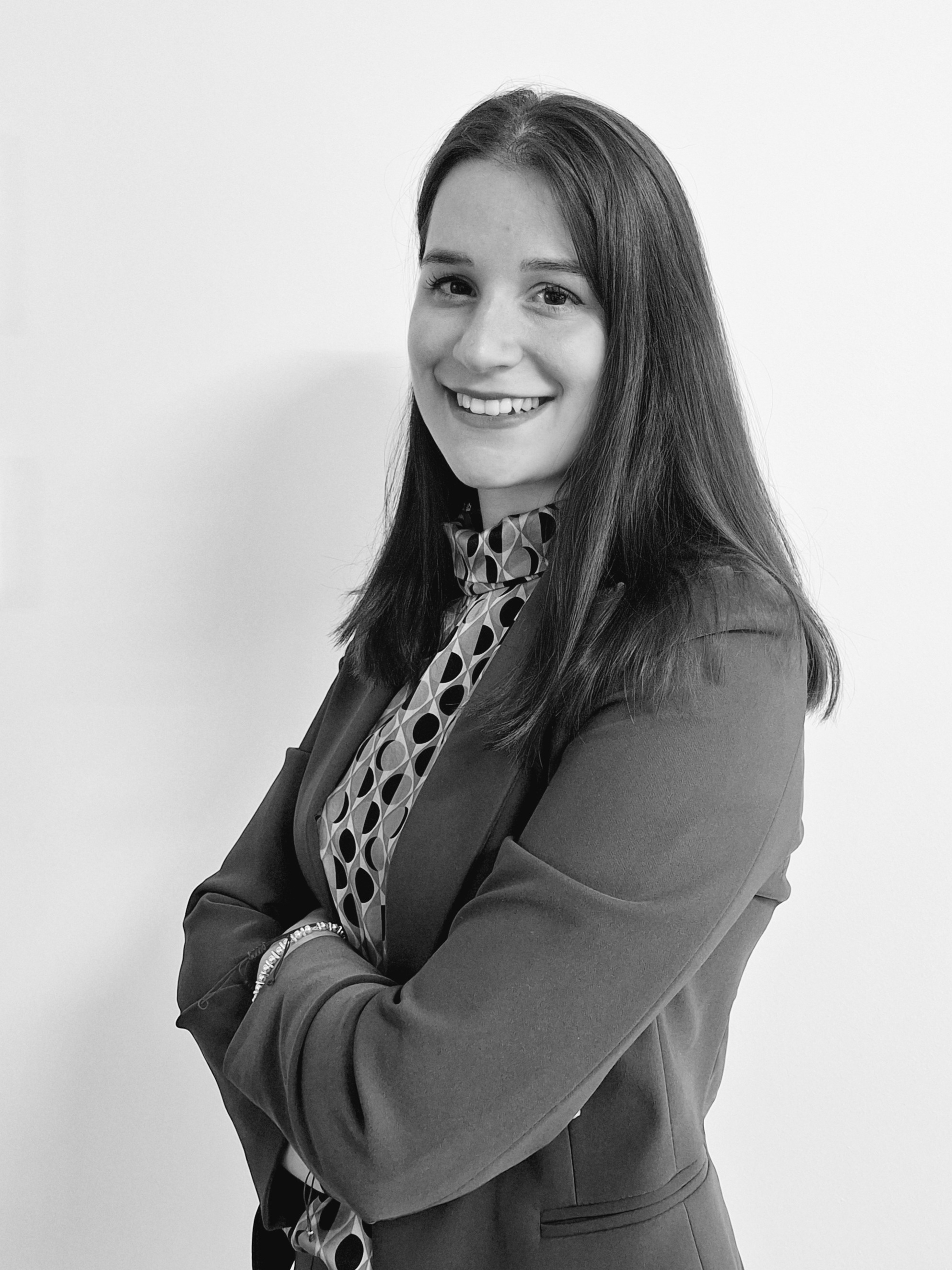}
  \end{wrapfigure}\par
  \noindent \textbf{Silvia Mura} received the M.Sc. and Ph.D. degrees cum laude in Telecommunications Engineering from Politecnico di Milano, Milan, Italy, in 2020 and 2023, respectively. She is currently an assistant professor at the Department of Electronics, Information, and Bioengineering at Politecnico di Milano. Her research interests encompass V2X communication systems, 
integrated sensing and communication systems, optimization and machine learning techniques, and design of physical layers for 6G wireless networks. \par

\vskip +1\baselineskip plus +1fil
\begin{wrapfigure}{l}{25mm} 
    \includegraphics[width=2.5cm,height=3.33cm,clip,keepaspectratio]{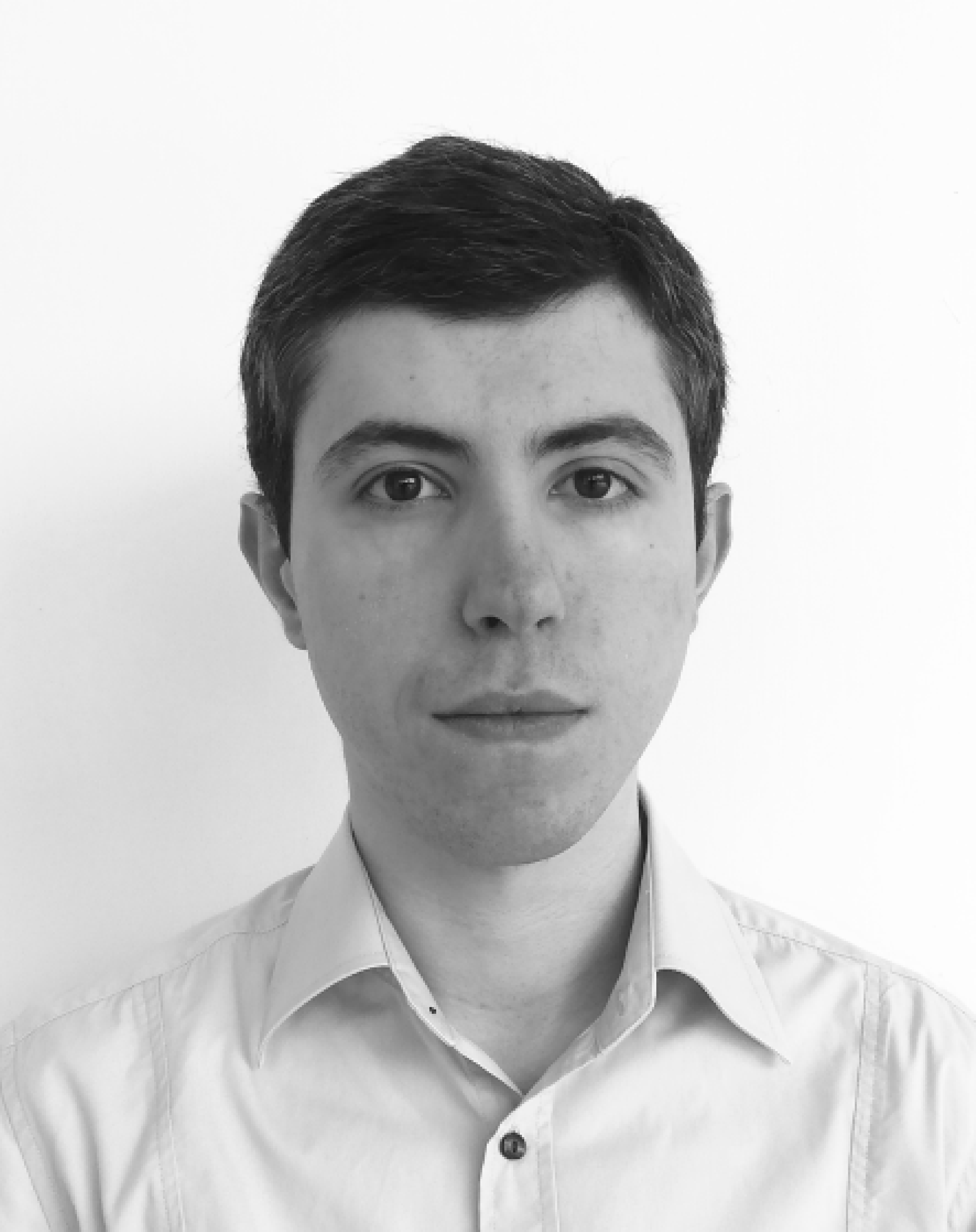}
  \end{wrapfigure}\par
  \noindent \textbf{Lorenzo Cazzella} received the B.Sc. degree in information engineering from Università del Salento, Lecce, Italy, in 2017, and the M.Sc. degree in computer science and engineering from Politecnico di Milano, Milan, Italy, in 2020, where he is currently pursuing the Ph.D. degree in information technology. During his master thesis, he has been involved in the Joint Laboratory between Huawei Technologies Italia, Milan, and Politecnico di Milano, where his research work nowadays concerns the study and development of deep learning solutions for channel estimation and integrated communication and radar sensing at the infrastructure. His research interests are currently focused on geometric deep learning and unsupervised learning techniques, with applications to wireless communication systems and radar systems. \par

\vskip +1\baselineskip plus +1fil
\begin{wrapfigure}{l}{25mm} 
    \includegraphics[width=2.5cm,height=3.33cm,clip,keepaspectratio]{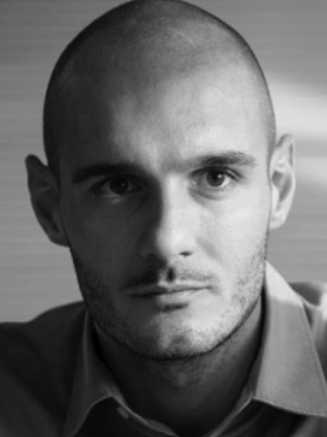}
  \end{wrapfigure}\par
  \noindent \textbf{Damiano Badini} received the bachelor’s degree (cum laude) in electrical engineering and the master’s degree (Hons.) in telecommunications engineering from Politecnico di Milano, Milan, Italy, in July 2013 and July 2016, respectively.,He joined the 5G Group of Huawei Technologies Italy Research Center, Milan, in November 2016, as an Algorithm and System Engineer. From October 2013 to October 2014, he was an undergraduate Researcher at Nokia Bell Laboratories, Holmdel, NJ, USA, where he worked on the realization of the first real-time multiple-input multiple-output (MIMO) system in multimode optical fibers. His main areas of interest lie in millimeter wave (mmWave) technologies and communications. \par

 \vskip +1\baselineskip plus +1fil
\begin{wrapfigure}{l}{25mm} 
    \includegraphics[width=2.5cm,height=3.33cm,clip,keepaspectratio]{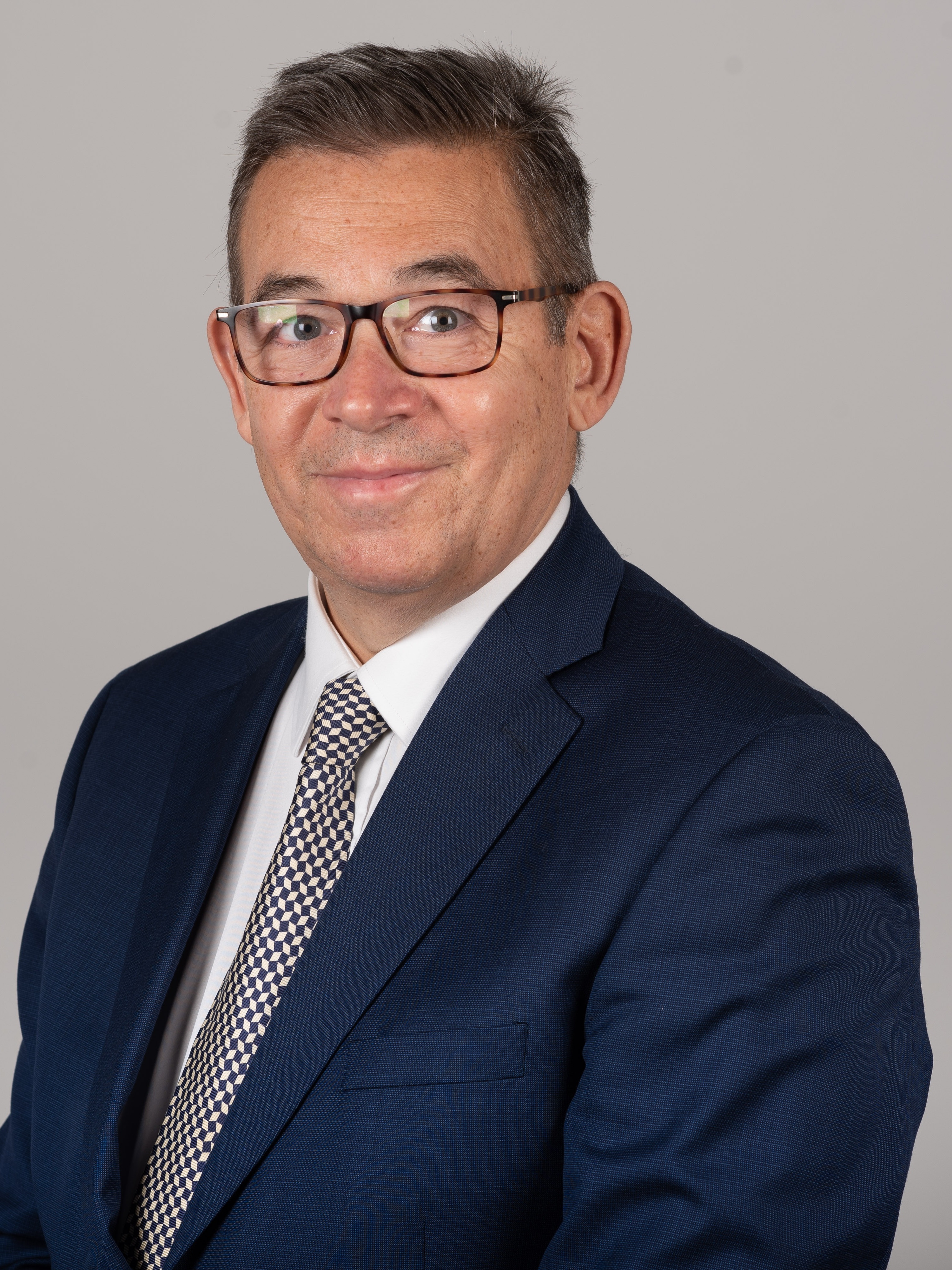}
  \end{wrapfigure}\par
  \noindent \textbf{Umberto Spagnolini} is a Professor of Statistical Signal Processing, the Director of the Joint Lab Huawei-Politecnico di Milano, and the Huawei Industry Chair of the Politecnico di Milano. His research in statistical signal processing covers remote sensing and communication systems with more than 380 papers on peer-reviewed journals/conferences and patents. He is the author of the book Statistical Signal Processing in Engineering (J. Wiley, 2017). He is the Technical Expert of standard-essential patents and IP. His research interests include mmW channel estimation and space-time processing for single/multiuser wireless communication systems, cooperative and distributed inference methods, including V2X systems, mmWave communication systems, parameter estimation/tracking, focusing and wavefield interpolation for remote sensing (UWB radar and oil exploration), and integrated communication and sensing. He was the recipient/co-recipient of the Best Paper Awards on Geophysical Signal Processing Methods (from EAGE), the Array Processing (ICASSP 2006), the Distributed Synchronization for Wireless Sensor Networks (SPAWC 2007, WRECOM 2007), the 6G Joint Communication and Sensing (JC\&S 2021), and the SAR Imaging for Automotive (MMS 2022). He served as a part of IEEE editorial boards as well as a member in technical program committees of several conferences for all the areas of interests.\par
\end{document}